\documentclass{elsart}
\usepackage{graphicx}
\usepackage{subfigure}

\begin{document}
\begin{frontmatter}
\title{Uncovering delayed patterns in noisy and irregularly sampled time series: an astronomy application}

\author[FI-UASLP,UBCS]{Juan C. Cuevas-Tello\thanksref{now}\corauthref{cor}},
\corauth[cor]{Corresponding author.}
\ead{cuevas@uaslp.mx}
\author[UBCS]{Peter Ti\v{n}o},
\author[UBPS]{Somak Raychaudhury},
\author[UBCS]{Xin Yao},
\author[HU]{Markus Harva}

\address[FI-UASLP]{Engineering Faculty, Autonomous University of San Luis Potosi, M\'{e}xico}

\address[UBCS]{School of Computer Science, University of Birmingham, UK}

\address[UBPS]{School of Physics and Astronomy, University of Birmingham, UK}

\address[HU]{Laboratory of Computer and Information Science, Helsinki University of Technology, Finland}

\thanks[now]{Present Address: Av Dr. Manuel Nava No.8, Zona Universitaria, San Luis Potos\'{i}, SLP, M\'{e}xico, ZC 78290}

\begin{abstract}
  We study the problem of estimating the time delay between two
  signals representing delayed, irregularly sampled and noisy versions
  of the same underlying pattern. We propose and demonstrate an
  evolutionary algorithm for the (hyper)parameter estimation of a
  kernel-based technique in the context of an astronomical problem,
  namely estimating the time delay between two gravitationally lensed
  signals from a distant quasar.  Mixed types (integer and real) are
  used to represent variables within the evolutionary algorithm. We
  test the algorithm on several artificial data sets, and also on real
  astronomical observations of quasar Q0957+561. By carrying out a
  statistical analysis of the results we present a detailed comparison
  of our method with the most popular methods for time delay
  estimation in astrophysics. Our method yields more accurate and more
  stable time delay estimates: for Q0957+561, we obtain 419.6 days
  between images A and B. Our methodology can be readily applied to
  current state-of-the-art optical monitoring data in astronomy, but
  can also be applied in other disciplines involving similar time
  series data.

\end{abstract}


\end{frontmatter}

\section{Introduction}

The 
estimation of {\it time delay}, the delay between arrival times of two
signals that originate from the same source but travel along different
paths to the observer, is a real-world problem in Astronomy. A time
series to be analysed could, for instance, represent the repeated
measurement, over many months or years, of the flux of radiation
(optical light or radio waves) from a very distant quasar, a very
bright source of light usually a few billion light-years away.  Some
of these quasars appear as a set of multiple nearby images on the sky,
due to the fact that the trajectory of light coming from the source
gets bent as it passes a massive galaxy on the way (the ``lens''),
and, as a result, the observer receives the light from various
directions, resulting in the detection of several images
\cite{Kochanek:2004:THC,Saha:2000:GL}. This
phenomenon is called gravitational lensing, and is a natural
consequence of a prediction of the General theory of Relativity, which
postulates that massive objects distort space-time and thus cause the
bending of trajectories of light rays passing near them.  Quasars are
variable sources, and the same sequence of variations
 is detected at different times in
the different images, according to the travel time along the various
paths. The time delay between the signals depends on the 
mass of the
lens, and thus it is the most direct method to measure the
distribution of matter in the Universe, which is often dark
\cite{Refsdal:1966:R66,Kochanek:2004:THC}.

In this scenario, the underlying pattern in time of emitted flux
intensities from a quasar gets delayed and corrupted by all kinds of
noise processes.  For example, astronomical time series are not only
corrupted by observational noise, but they are also typically
irregularly sampled with possibly large observational gaps (missing
data)
\cite{Ovaldsen:2003:NAP,Pindor:2005:DGL,Oguri:2007:STT,Inada:2008:IND}. This
is due to practical limitations of observation such as equipment
availability, weather conditions, the brightness of the moon, among
many other factors \cite{Eigenbrod:2005:TCM}. Over a hundred systems
of lensed quasars are currently known\footnote{A growing list of
  multiply-imaged gravitationally lensed quasars can be found at
  http://cfa-www.harvard.edu/castles.}, and about 10 of these have
been monitored for long periods, and in some of these cases, the
measurement of a time delay has been claimed.  Here we focus on
Q0957+561, the first multiply-imaged quasar to be discovered
\cite{Walsh:1979}. This source, which has a pair of images (here
referred to as A and B), has been monitored for over twenty years, and
despite numerous claims, a universally agreed value for the time delay
in this system has not emerged \cite{Kundic:1997:ARD,Cuevas:2006:HAT}.

In an earlier paper, we presented an analysis of repeated radio
observations, along with simulated data generated according to the
properties of these observations
\cite{Cuevas:2006:HAT}, to show that a kernel-based approach can
improve upon the currently popular methods of estimating time delays
from real astronomical data. The more common form of observations,
however, employs optical telescopes for monitoring known
multiply-imaged sources, and these observations have inherent problems
that require the modification of our previous approach. Here we
present a largely modified approach that outperforms  on optical datasets our previous appraoch, as well as alternative approaches in use in astrophysics.

Here we introduce a novel 
evolutionary algorithm (EA) to estimate the parameters of a
model-based method for time delay estimation. The EA uses, as a
fitness function, the mean squared error (MSE$_{CV}$) given by
cross-validation on observed data, and also performs a novel regularisation procedure based on singular value decomposition (SVD).
Our population is also represented by mixed types, integers and reals. 

The contribution of this paper is in several directions: i) an evolutionary algorithm has been introduced to form a novel
hybridisation with our kernel method, ii) a principled automatic method has been proposed to estimate
the time delay, kernel width, and SVD regularisation parameters, iii) the application of EA driven by a model based
formulation to a real-world problem, and iv) we carefully study statistical significance of the results on different data. 

Our EA is an evolutionary optimisation technique in presence of uncertainties \cite{Jin:2005:EOU} 
and missing data with mixed representation -- through two linked populations,
each devoted to one particular data type. 
The parameters to optimise come from a kernel machine. We do 
parameter optimisation and 
model selection at the same time.
This approach can be applied to other problems, not only time series from gravitational lensing.
For instance, the missing data problems cover those cases where instrumental equipment fails, observations are 
incorrectly recorded, sociological factors are involved, etc. Therefore, the data are unevenly sampled,
which restricts the use of Fourier analysis \cite{Press:2002:book}(\S13.8). Problems with noisy and missing data
occur in almost all sciences, where the data availability is influenced by what is easy or feasible to collect (e.g., see \cite{Cook:2004:WMDP,Bridewell:2006:LPM}). 

We compare the performance of our
EA in several ways:

\begin{enumerate}
\item
 The
performance of our method is assessed against that of two of the most
popular methods in the astrophysical literature \cite{Vuissoz:2008:COSMOGRAIL,Fohlmeister:2008:TRP,Ovaldsen:2003:NAP,Colley:2003:ATC,Haarsma:1999:TRW}, i.e., 
{\bf (a)} the Dispersion spectra method
\cite{Pelt:1994:TDC,Pelt:1996:TLC,Pelt:1998:EMT,Cuevas:2006:HAT} and
{\bf (b)} a scheme based on the structure function of the
intrinsic variability of the source, here referred to as the PRH
method \cite{Press:1992:TTD}. 

\item
Because the true time delay of observed fluxes from quasars is not known,
we assess the performance of algorithms in a controlled series
of experiments, where artificially generated data with known delays are used.
We employ three kinds of artificial data
sets: large scale data \cite{Cuevas:2006:HAT}, PRH data
\cite{Press:1992:TTD,Cuevas:2006:HAT} and Wiener data (as outlined in
\cite{Harva:2006:IEEE,HarvaR08}). 

\item
To justify our EA, an analogous non-evolutionary model-based approach (K-V) is also
employed in this paper. 

\end{enumerate}

Our statistical analysis shows that the results from our EA are 
more accurate and significant than state-of-the-art methods.
We use our
EA as well as a (1+1)-ES algorithm \cite{Rowe:2004:ES} on actual astronomical observations, where
the twin images were observed over several years with optical
telescopes \cite{Kundic:1997:ARD}.

The remainder of this paper is organised as follows: the data under
analysis is described in \S\ref{data-section}. The kernel approach is
outlined in \S\ref{kernel-section}, and the EA is presented in \S
\ref{evo-section}. 
The results and our conclusions are in \S\ref{results-section} and \S\ref{conclusion-section} respectively. Finally, our future work is presented in \S \ref{future-section}.


\section{Data}\label{data-section}
\subsection{Optical Data}\label{optical-section}
In this paper, we use optical observations\footnote{Astronomers
observe quasars at other wavelengths as well, e.g., with
radio telescopes \cite{Haarsma:1999:TRW}.} of the two images of the
quasar Q0957+561, from a monitoring program at the Apache Point
Observatory, New Mexico, USA \cite{Kundic:1997:ARD}. This data set has
97 observations, where, in each observation, fluxes are measured
of all the multiple images of the source, 
in the $g$-band (a standard yellow-green filter),
from December 1994 to July 1996.  

\begin{figure}
\centering
\includegraphics[width=3.5in]{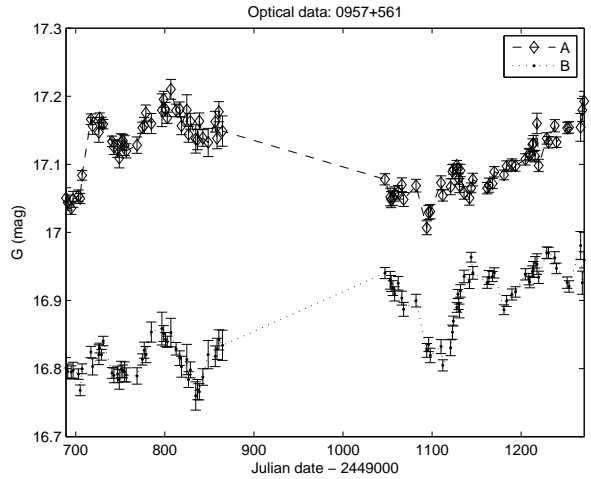}
\caption{Observations of the brightness of
the doubly-imaged quasar Q0957+561, in the $g$-band,
as a function of time (Top: Image A;
Bottom: Image B, see Table~\ref{optical-table}).
The time is measured in days (Julian days--2,449,000 days).
}
\label{optical-fig}
\end{figure}

The observed time series (here called {\it light curves}) are given
in Table~\ref{optical-table}, where the Time column, representing the
time 
of observation (note that it is irregularly sampled), is given in Julian days (JD, defined as the
number of days since Noon GMT on January 1, 4713 BC).
The fluxes observed from images A and B are given in the astronomical
unit of magnitude (mag $m$), defined as $m=-2.5\,\log_{10}\,f$, where
$f$ is the flux measured when observed through a green filter\footnote{This data set is available online
\cite{Kundic:1997:ARD}.} ($g$-band). In Fig.~\ref{optical-fig}, the time series are
shown. The measurement errors, which are standard deviations
(std) of the flux measurement, are given in the Table~\ref{optical-table} as Error A and Error
B; these are the error bars. The source was monitored nightly, but many
observations were missed due to cloudy weather and telescope
scheduling.  The big gap in Fig.~\ref{optical-fig} is an intentional
gap in the nightly monitoring, since a delay of about 400 days, the pattern, was
known `a priori' -- monitoring programs on this quasar started in 1979. Therefore, the peak in the light curve of
image~A, between 700 and 800 days, corresponds to the peak in that of
image~B between 1,100 and 1,200 days.

\begin{table}
\renewcommand{\arraystretch}{1.3}
\caption{Optical data: Q0957+561 observed in the $g$-band, from \cite{Kundic:1997:ARD}}
\label{optical-table}
 \begin{tabular}{c c c c c}
 \hline
  Time & Image A &  Error A & Image B & Error B \\
  (days) & (mag) &   & (mag) &  \\
 \hline
   689.009 & 16.9505 &  0.0152 & 16.8010 &  0.0152 \\
   691.007 & 16.9439 &  0.0111 & 16.7957 &  0.0111 \\
   695.001 & 16.9356 &  0.0090 & 16.7949 &  0.0090 \\
     ...   &  ...    &  ...    &  ...    &  ...    \\
  1253.672 & 17.0544 &  0.0084 & 16.9206 &  0.0084 \\
  1266.665 & 17.0544 &  0.0205 & 16.9808 &  0.0205 \\
  1268.642 & 17.0798 &  0.0170 & 16.9261 &  0.0170 \\
  1270.652 &  17.0928 &  0.0145 &16.9597 &  0.0119 \\
\hline
\end{tabular}
\end{table}

\subsection{Artificial Data}\label{artificial-section}
Since the definite time delay for the Q0957+561 is unknown, one cannot
test the accuracy of methods through real data. Therefore, many
attempts have been made to generate synthetic data in order to test
the performance of methods
(e.g. \cite{Press:1992:TTD,Pijpers:1997:TDT,Burud:2001:ANA,Eigenbrod:2005:TCM,Harva:2006:IEEE,HarvaR08}). Below
we describe three kinds of artificial data sets that we have used,
representing the major classes of data sets used by others: large
scale data, PRH data and Wiener data.


\subsubsection{Large Scale Data}\label{large-scale-data-section}

\begin{figure}[h]
\centering
\begin{tabular}{c}
\subfigure[]{\includegraphics[width=3.0in]{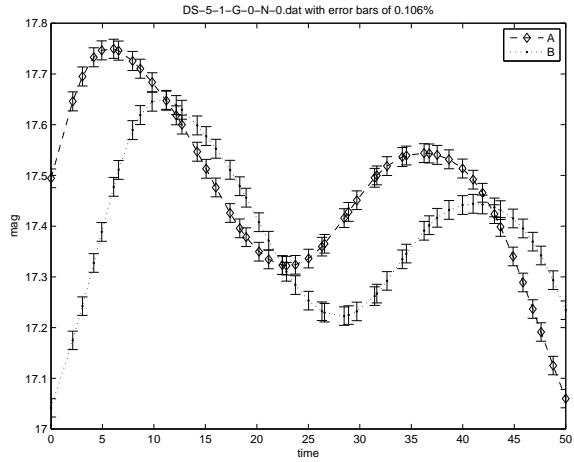}} \\
\subfigure[]{\includegraphics[width=3.0in]{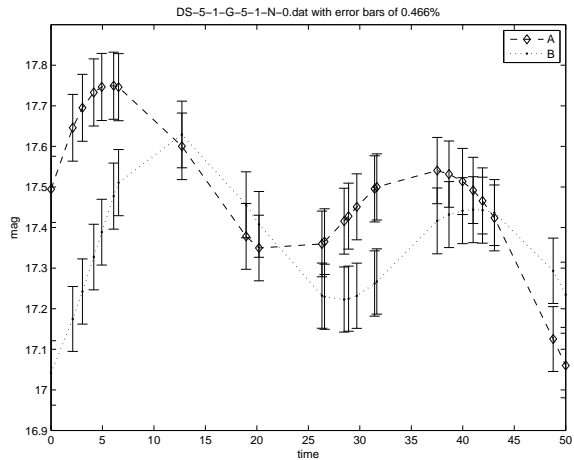}} 
\end{tabular}
\caption{The simulated Large Scale Data sets, as outlined in 
Table~\ref{large-data-table}. {\bf (a)} The first underlying function
(DS-5-1) without noise and no gaps. Error bars represent 0.106\% of
mag.{\bf (b)} This data set corresponds to the same underlying
function (DS-5-1), without noise, and the gap size is five (first
realisation). Error bars represent 0.466\% of flux.}
\label{ds-5-fig}
\end{figure}

In this data set (DS-5), the true time delay is 5~days
\cite{Cuevas:2006:ECML}, and the true offset in brightness between 
image~A and image~B is $M=0.1$~mag.  The intention here is to simulate
optical observations as in Ovaldsen et
al. \cite{Ovaldsen:2003:NAP}. We employ only the first five underlying
functions\footnote{Plots are available at
http://www.cs.bham.ac.uk/$\sim$jcc/artificial-optical/}. These data
sets are irregularly sampled with three levels of noise and gaps of
different size as shown in Table~\ref{large-data-table}; for more
details see \cite{Cuevas:2006:HAT,Cuevas:2006:ECML}. We use 50
realisations per level of noise only. Consequently, this yields 38,505
data sets (see Table~\ref{large-data-table}), with 50 samples each,
of which two 
are shown in Fig.~\ref{ds-5-fig}. 

\begin{table}
\caption{Simulated Large Scale Data sets}
\label{large-data-table}
  \begin{tabular}{ c l l l l l l }  
  \hline 
                &\multicolumn{5}{c}{Gap size} & \\ 
  \textbf{Noise}&\textbf{0}&\textbf{1}&\textbf{2}&\textbf{3}&\textbf{4}&\textbf{5} \\  \hline
      0\%  & 1   & 10   & 10   & 10   & 10   & 10    \\
  0.036\%  & 50 & 500 & 500 & 500 & 500 & 500  \\
  0.106\%  & 50 & 500 & 500 & 500 & 500 & 500  \\
  0.466\%  & 50 & 500 & 500 & 500 & 500 & 500  \\ \hline
   Sub-Total & 151 & 1510 & 1510 & 1510 & 1510 & 1510 \\ \hline 
   \multicolumn{7}{l}{Total = 7,701 data sets per underlying function.} \\
   \multicolumn{7}{l}{5 underlying functions yield 38,505 data sets.} \\
 \end{tabular}
\end{table}

\subsubsection{PRH Data}
\begin{figure}[h]
\centering
\begin{tabular}{c c}
\subfigure[]{\includegraphics[width=3.5in]{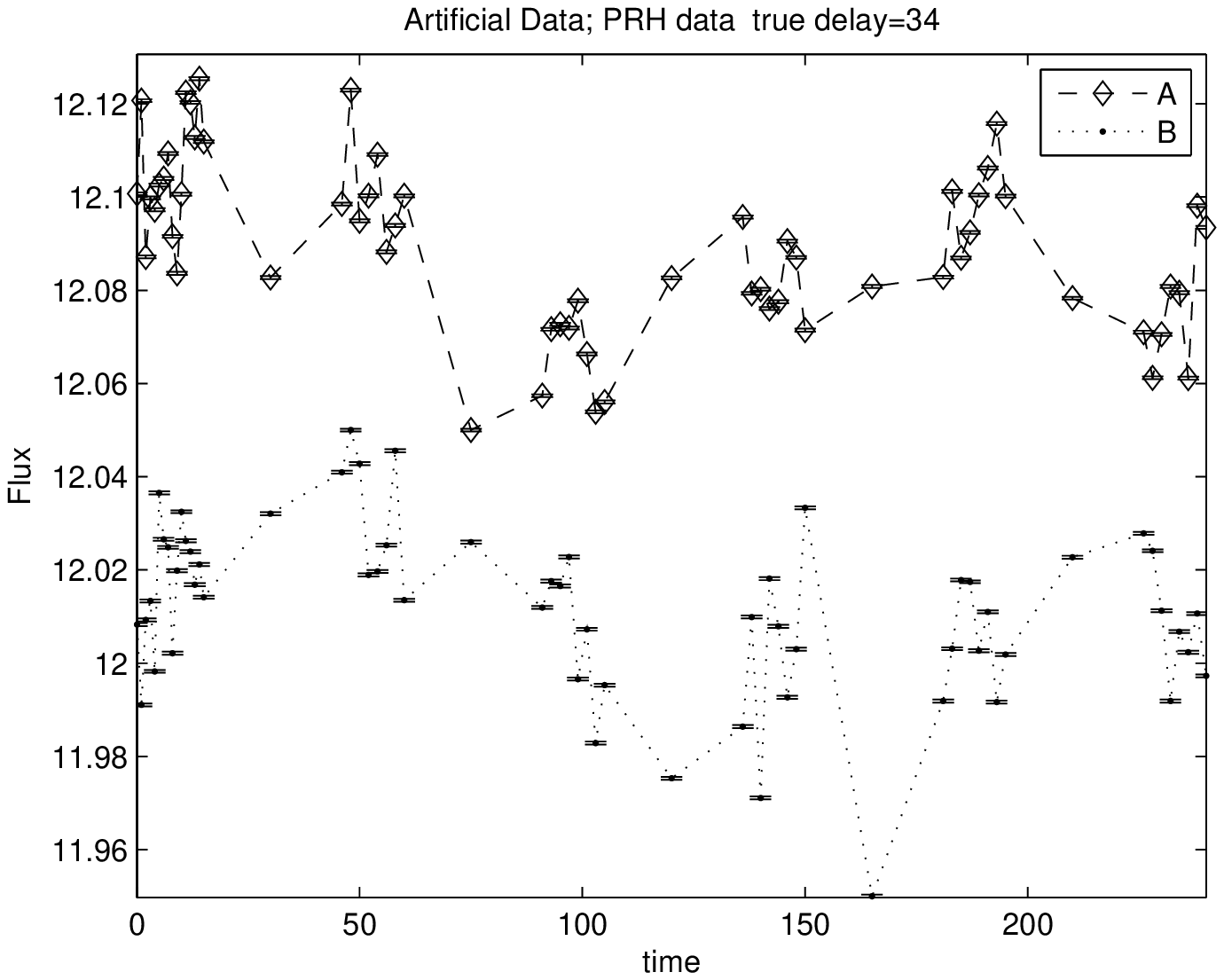}} \\
\subfigure[]{\includegraphics[width=3.5in]{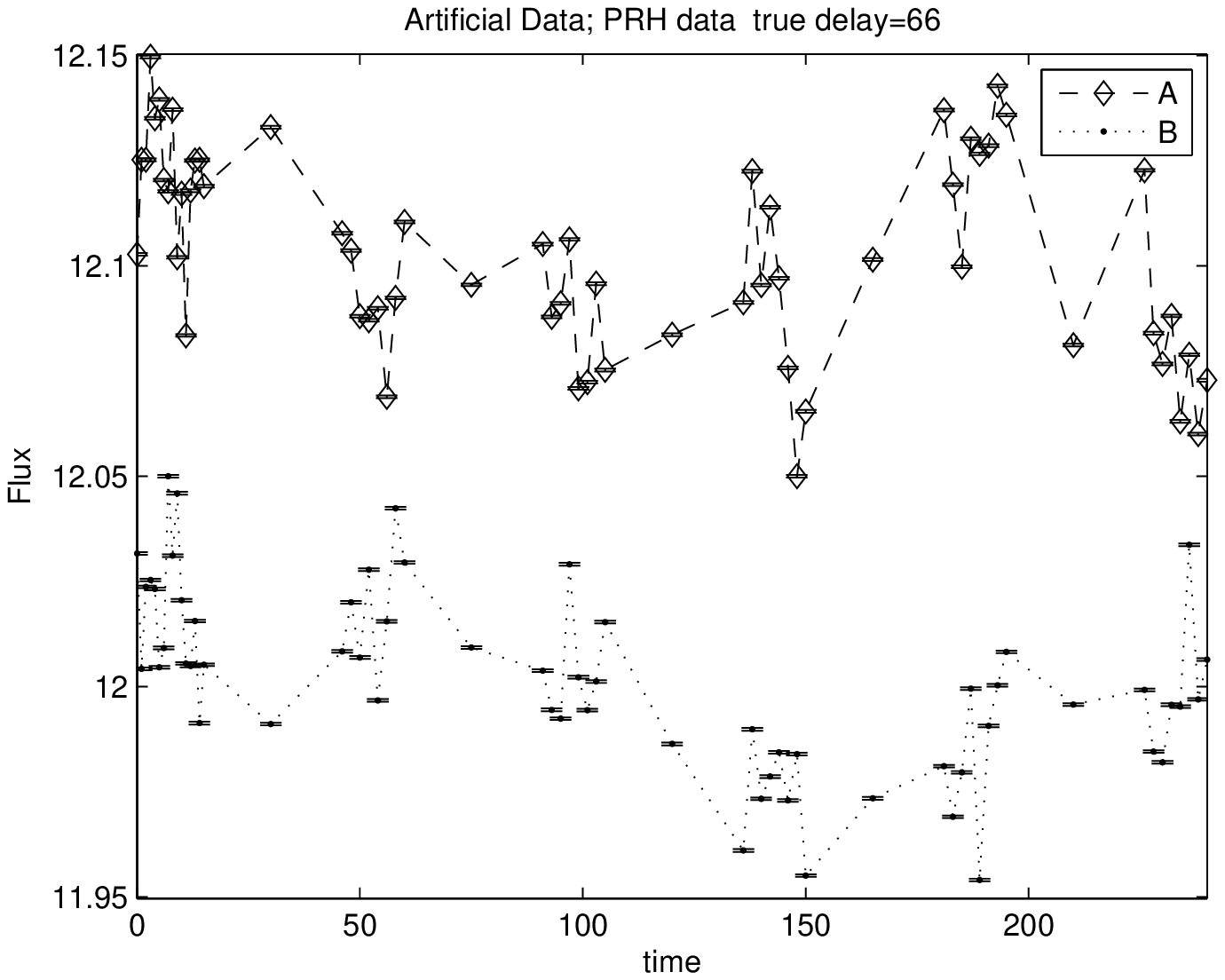}} 
\end{tabular}
\caption{Examples of the simulated PRH Data. 
The error bars represent a variance of $1 \times 10^{-7}$. {\bf (a)}
This is a realisation for a true delay of 34~days. Image A has been
shifted upwards by 0.08 for visualisation. {\bf (b)} In this
realisation, the true delay is 66 days. 
Image A has been shifted upwards by
0.1 for visualisation. }
\label{PRH-data-fig}
\end{figure}

These data sets are generated by Gaussian processes, following
\cite{Press:1992:TTD}, with a fixed covariance matrix given by a
structure function according to Pindor \cite{Pindor:2005:DGL}. The
variance representing the measurement errors is $1 \times
10^{-7}$. They are highly sampled with periodic gaps
\cite{Eigenbrod:2005:TCM}, simulating a monitoring campaign of eight
months; yielding 61 samples per time series. There are seven true
delays and 100 realisations for each value of true delay
\cite{Cuevas:2006:HAT}. Two plots are shown in Fig.~\ref{PRH-data-fig}.

\subsubsection{Wiener Data}
\begin{figure}[h]
\centering
\begin{tabular}{c}
\subfigure[]{\includegraphics[width=3.5in]{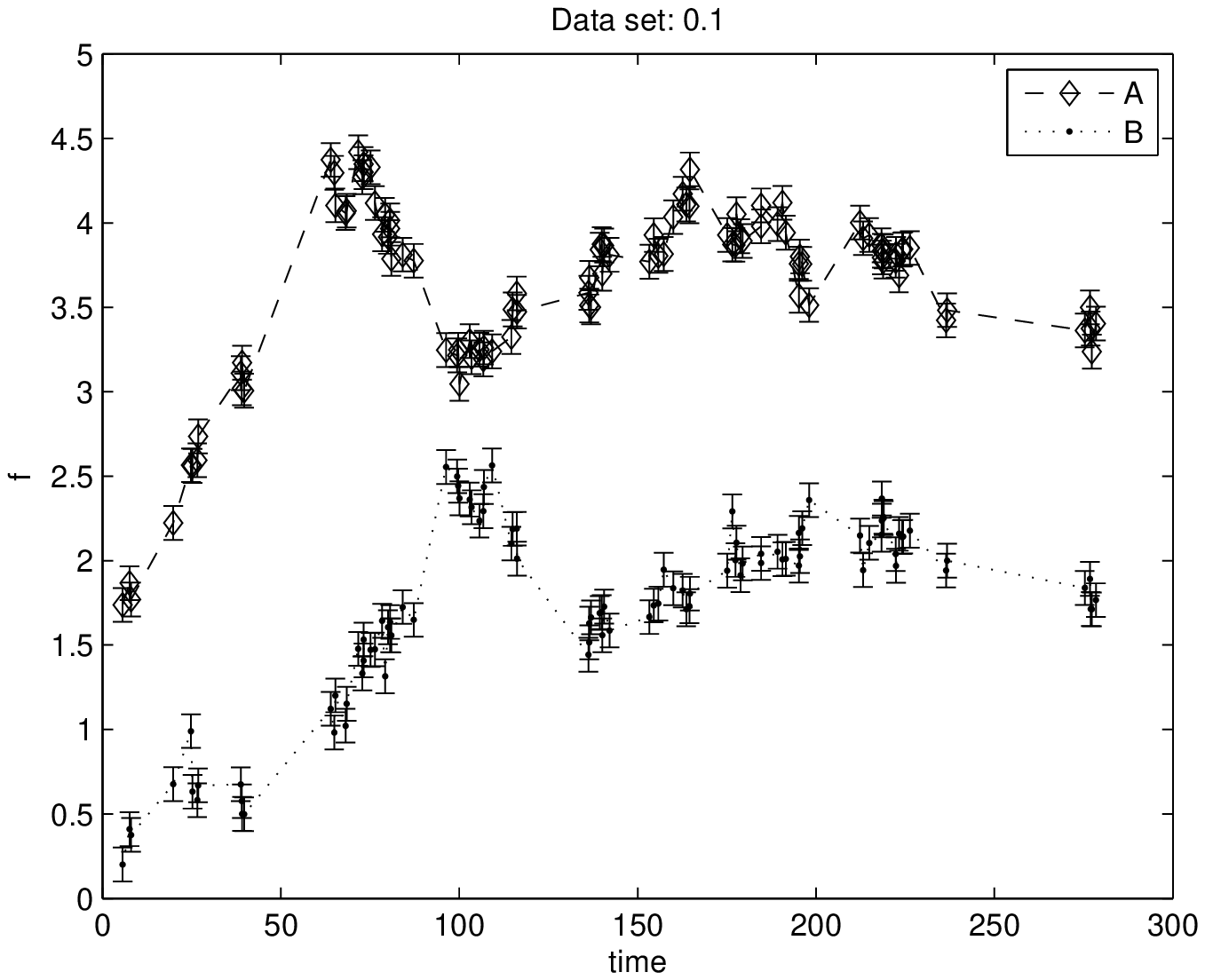}} \\
\subfigure[]{\includegraphics[width=3.5in]{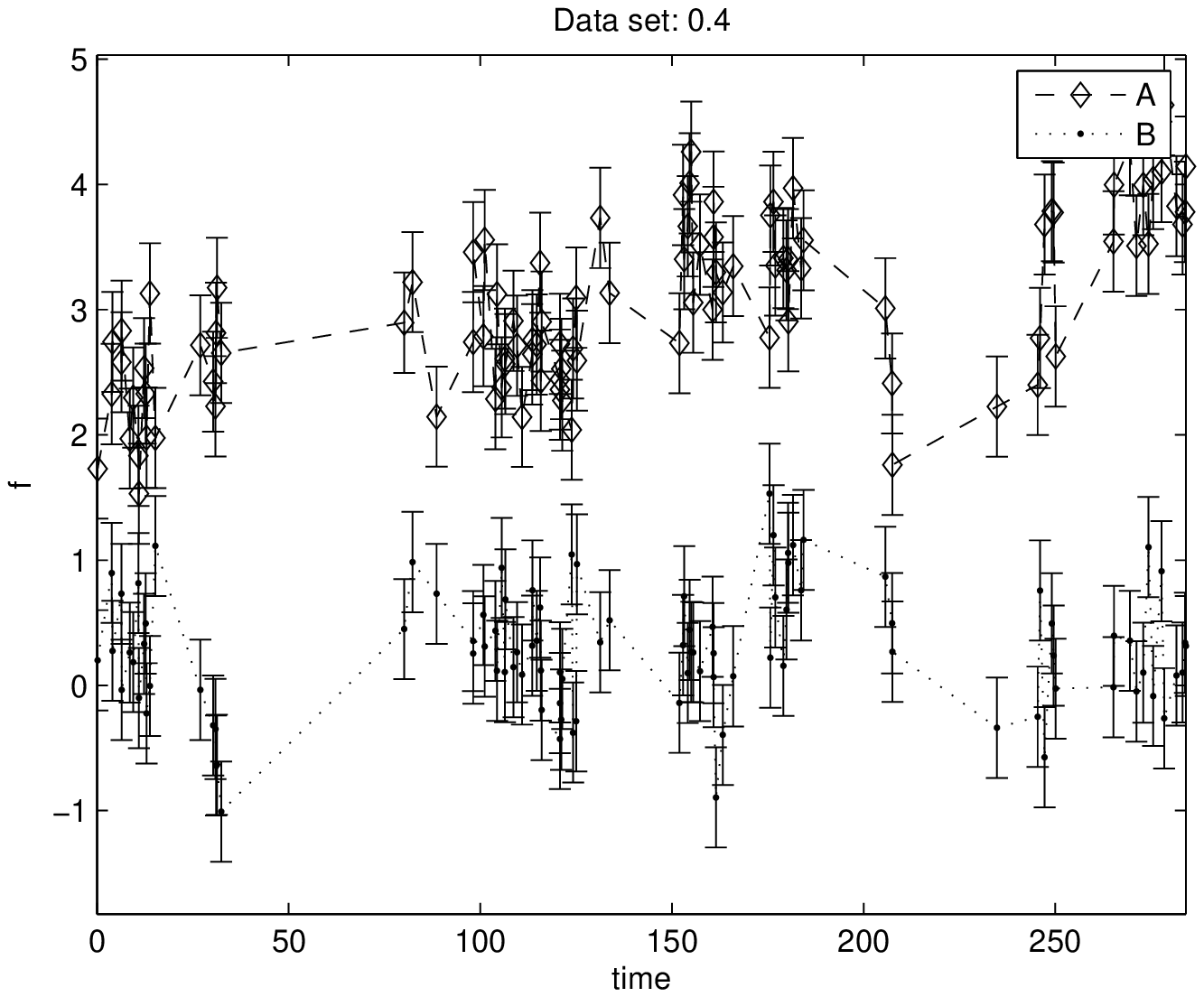}} 
\end{tabular}
\caption{Simulated Wiener Data. 
{\bf (a)} The first realisation of data sets with noise of variance
$0.1^2$. Image A has been shifted upwards by a factor 1.5 for
visualisation. {\bf
(b)} The first realisation of data sets 
with noise of variance
$0.4^2$. Image A has been shifted
upwards by 2.9 for visualisation. In each case,
error bars represent the standard deviation.}
\label{harva-data-fig}
\end{figure}

These data sets, generated by a Bayesian model \cite{Harva:2006:IEEE},
simulate three levels of noise with 225 data sets per level of noise,
where each level of noise represents the variance: $0.1^2$, $0.2^2$
and $0.4^2$. The data are irregularly sampled and the true time delay
in all cases is 35 days. Some examples are shown in
Fig.~\ref{harva-data-fig}. Each time series has 100 samples.

\section{Kernel Approach}\label{kernel-section}

In previous papers, 
\cite{Cuevas:2006:HAT,Cuevas:2006:ECML}, 
we have introduced a kernel-based approach, and we would refer the
reader to these papers for further detail, since not all derivations
will be repeated here at the same level of detail.  The aim of this
section is to come up with the parameters to evolve in \S
\ref{evo-section}.

We model a pair of time series, obtained by monitoring the brightness
of images A and B (see \S \ref{data-section}), as 
\begin{equation}
  \label{data}
  \begin{array}{ll}
      x_{A}(t_{i}) & = h_{A}(t_{i})+\varepsilon_{A}(t_{i}) \\
      x_{B}(t_{i}) & = h_{B}(t_{i}) \ominus M + \varepsilon_{B}(t_{i}),
  \end{array}
 \end{equation}
\noindent 
where $\ominus=\{\times,-\}$ denotes either multiplication or
subtraction, so $M$ is either a ratio (used in radio observations,
where brightness is quoted in flux units) or an offset
between the two images
(as in optical observations, where brightness in represented
in logarithmic units). We use the latter option here. 
Values of the independent variable $t_{i},i=1,2,...,n$ represent
discrete observation times. The observation errors
$\varepsilon_{A}(t_{i})$ and $\varepsilon_{B}(t_{i})$ are modelled as
zero-mean Normal distributions
\begin{equation}
  \label{error-model}
   N(0,\sigma_{A}(t_{i})) \ \ \hbox{and} \ \ N(0,\sigma_{B}(t_{i})),
\end{equation}
\noindent 
respectively, where 
$\sigma_A(t_i)$ and $\sigma_B(t_i)$ are standard deviations. 
Now,
\begin{equation}
  \label{hA}
    h_{A}(t_{i})=\sum_{j=1}^N\alpha_{j}K(c_{j},t_{i})
\end{equation}
\noindent 
is the ``underlying'' light curve that underpins image~A, whereas
\begin{equation}
  \label{hB}
    h_{B}(t_{i})=\sum_{j=1}^N\alpha_{j}K(c_{j}+\Delta,t_{i})
\end{equation}
is a time-delayed (by $\Delta$) version of $h_{A}(t_{i})$
underpinning image~B.

The functions $h_{A}$ and $h_B$ are formulated within the generalised
linear regression framework
\cite{Hastie:2001:book,Shawe-Taylor:2004:KM}. Each function is a
linear superposition of $N$ kernels $K(\cdot,\cdot)$ centred at either
$c_j$, $j=1,2,...,N$ (function $f_A$), or $c_j+\Delta$, $j=1,2,...,N$
(function $f_B$).  We use Gaussian kernels of width $\omega_{c}$: \
for $c,t \in \Re$,
\begin{equation}
  \label{kernel}
    K(c,t)=\exp\, {\frac{-|t-c|^{2}}{\omega_{c}^{2}}}.
    \end{equation}
The kernel width $\omega_{c}>0$ determines the `degree of smoothness'
of the models $h_A$ and $h_B$ . We position kernels at the position
of each  observation, implying $N=n$. 
The width $ \omega_j \equiv \omega_{c}$ is determined through the 
$k$ nearest neighbours of $c_j$ (equal to $t_j$) as 
\begin{equation}
  \label{k-equation}
\omega_{j}=\sum_{d=1}^{k}(t_{j} - t_{j-d})+(t_{j+d} - t_{j})= \sum_{d=1}^{k}(t_{j+d} - t_{j-d}).
\end{equation}

The weights $\vec{\alpha}$ (\ref{hA})-(\ref{hB}) are obtained as follows \cite{Cuevas:2006:HAT}:
\begin{equation}
  \label{alphas}
   \vec{K} \vec{\alpha} = \vec{x},
\end{equation}
\noindent 
where
$\vec{\alpha} = (\alpha_1, \alpha_2, ..., \alpha_N)^T$,
\begin{equation}
\label{alpha-kernel_data}
  \vec{K} =  \left[
              \begin{array}{c}
                K_{A}(\cdot,\cdot)\\ \\
                K_{B}(\cdot,\cdot)
              \end{array}
             \right],
\  \ \ \ \ \ \ \
\vec{x} =
            \left[
              \begin{array}{c}
                x_{A}(\cdot)/\sigma_{A}(\cdot) \\ \\
                x_{B}(\cdot)/\sigma_{B}(\cdot)
              \end{array}
             \right],
\end{equation}
and the kernels $K_A(\cdot,\cdot)$, $K_B(\cdot,\cdot)$ have the form:
\begin{equation}
  \label{kA-kB}
    K_{A}(c,t) = \frac{K(c,t)}{\sigma_{A}(t)}, \ \ \ \ \ \ \
    K_{B}(c,t) = \frac{M \ominus  K(c+\Delta,t)}{\sigma_{B}(t)}.
\end{equation}
\noindent 
Hence,
\begin{equation}
  \label{inverse}
   \vec{\alpha} = \vec{K}^{+}\vec{x}.
\end{equation}

Our aim is to estimate the time delay $\Delta$ between the temporal
light curves corresponding to images A and B. Typically, $\Delta$ is
estimated by a set of trial time delays in the range
[$\Delta_{min}$, $\Delta_{max}$] with a specific measurement of
goodness of fit \cite{Cuevas:2006:HAT}. In Eq. \ref{inverse}, the superscript ``+'' represents a pseudoinverse of a matrix, the pseudoinverse rather than the inverse is required because the matrix is not squared, that is, an over-determined system is involved \cite{Press:2002:book}.

Finally, the parameters are 
the time delay $\Delta$ [as given in Eqs.~(\ref{hA}) \& (\ref{hB})], 
the variable width $k$
[as in Eq.~(\ref{k-equation})], and the regularisation parameter 
$\theta$ (see below).

\subsection{Regularisation}\label{regularisation-section}
In practice, the matrix $\vec{K}$ (\ref{alpha-kernel_data}) may be
singular because $\vec{K}$ is an over-determined system, and noisy time
series (\ref{error-model}) are involved. We therefore regularise the
inversion in (\ref{inverse}) through singular value decomposition
(SVD) \cite{Cuevas:2006:HAT}. To avoid singularity, the most
straightforward method is to find a threshold $\lambda$ for singular
values \cite{Press:2002:book,Golub:1989:book}. This means that
the  singular values
less than $\lambda$ are set to zero, following which $\vec{K}^{+}$
(\ref{inverse}) is obtained through SVD \cite{Press:2002:book}.

\begin{figure*}[t!]
\centering
\begin{tabular}{c c}
\subfigure[]{\includegraphics[width=2.5in]{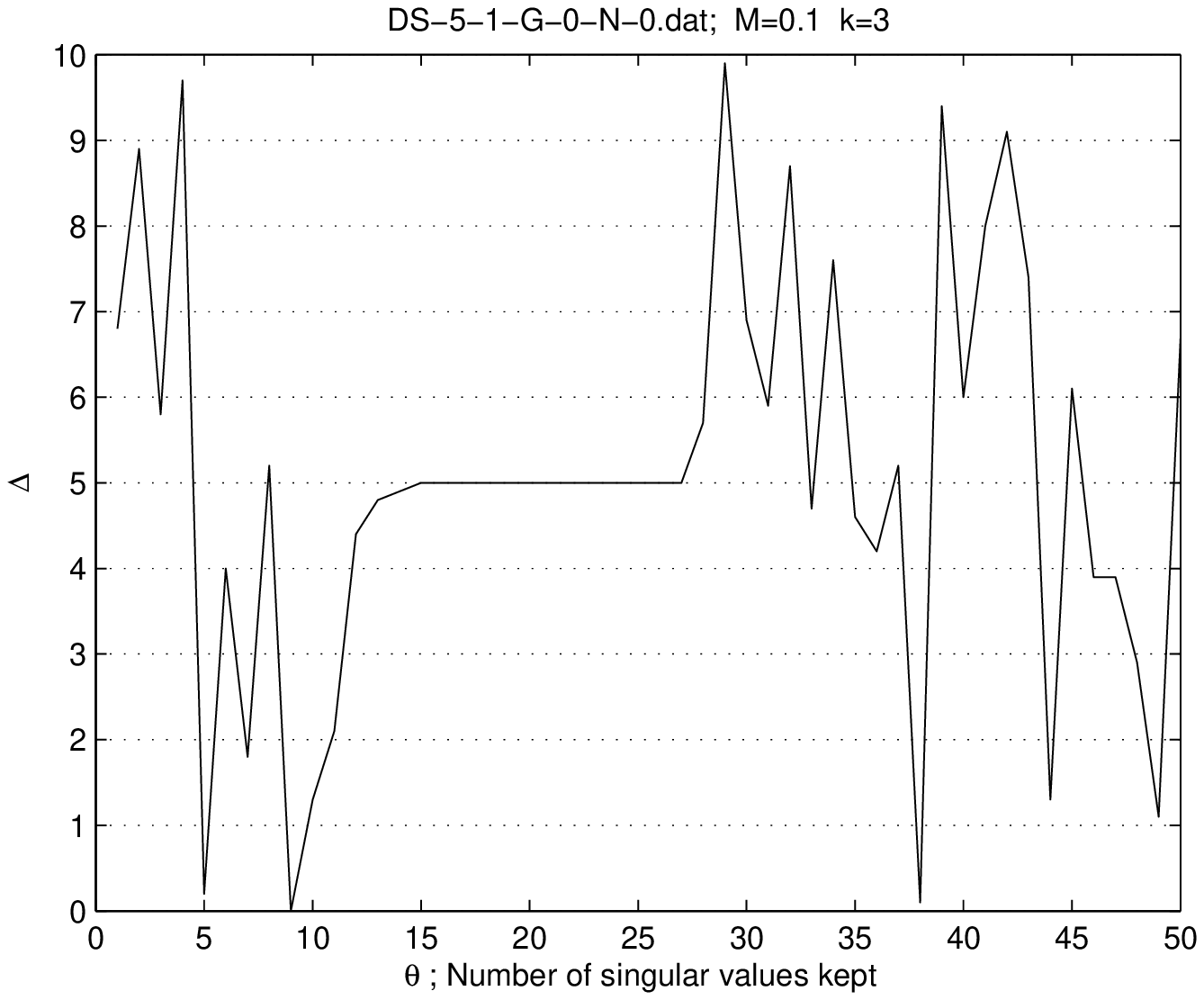}} &
\subfigure[]{\includegraphics[width=2.5in]{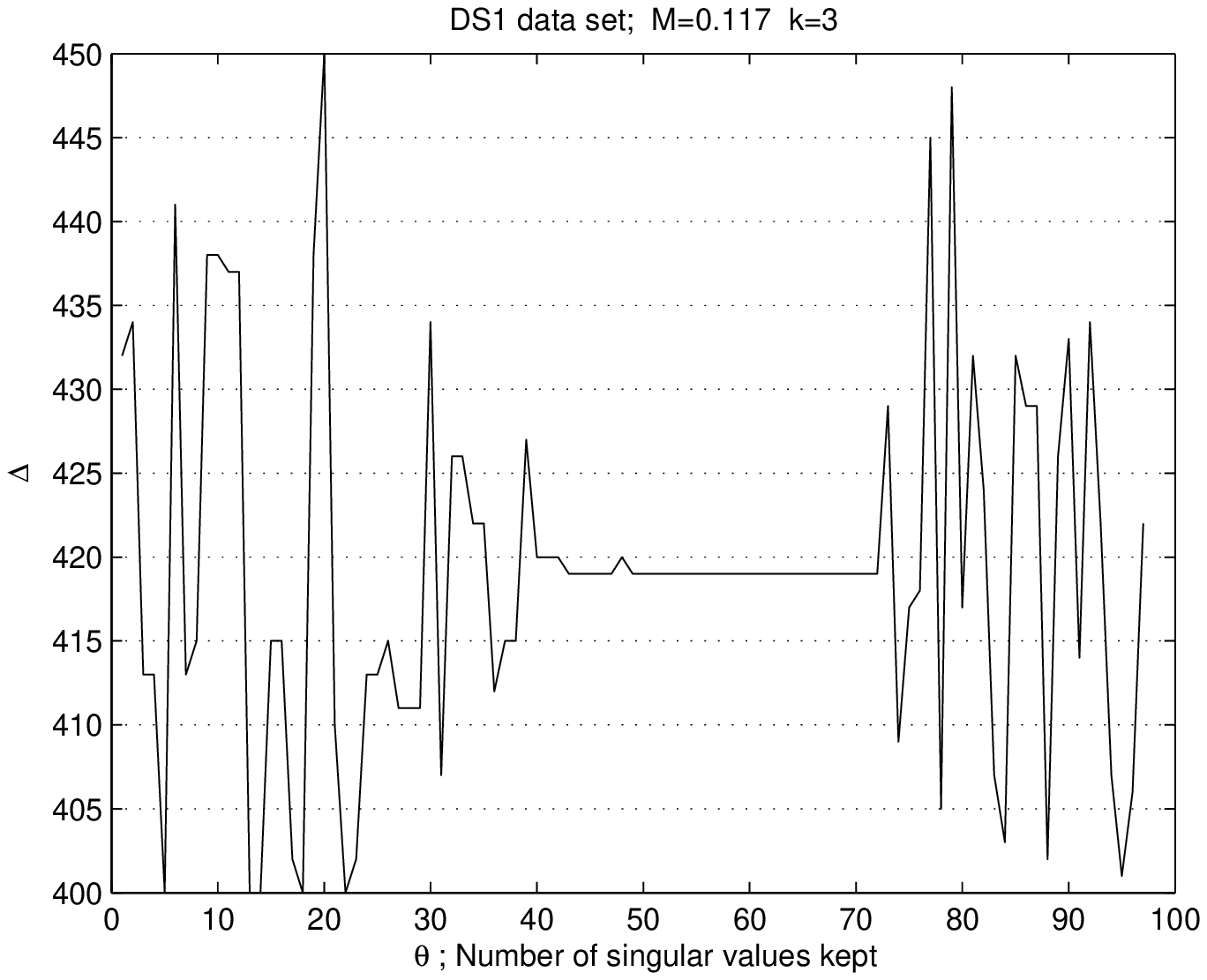}} 
\end{tabular}
\caption{Patterns. 
In each relation ($\Delta$, $\theta$), the best time delay has been
plotted, i.e., the best $\Delta$ ({\it y}-axis) versus the number of
singular values $\theta$ set to zero ({\it x}-axis). The best time
delay is found through log-likelihood \cite{Cuevas:2006:HAT} by
evaluating time delay trials in a given range. {\bf (a)}
DS-5-1-G-0-N-0. This data set has no noise and no gaps;
$\Delta=[0,10]$ with increments of 0.1; $M$ is set to its true value
$M=0.1$ (for more details concerning these data, see 
\S\ref{large-scale-data-section}). The pattern is at $\theta= [5, 27]$,
where $\Delta = 5$ (true value).  {\bf (b)} Actual $g$-band
optical observations of quasar Q0957+561,
$\Delta=[400, 450]$ with unitary increments. The data set is
as in \S\ref{optical-section}; $M$ was set to 0.117
\cite{Kundic:1997:ARD} and $k=3$ (\ref{k-equation}). The pattern is at
$\theta = [49, 72]$, where $\Delta = 419$ (see Tables~\ref{result1-table} 
and \ref{result2-table}, and \S\ref{results-section}). }
\label{pattern-fig}
\end{figure*}

In other words, $\lambda$ tells us how many singular values to set to
zero. Hence, for a given $\Delta$, the number of singular values to
keep may vary. We illustrate this through Fig.~\ref{pattern-fig},
representing artificial and optical data, where $\theta$ is the number
of singular values to set to zero. One can see a well defined pattern
in the range $\theta= [15, 27]$ ($\Delta = 5$) in
Fig.~\ref{pattern-fig}a, and $\theta= [49, 72]$ ($\Delta = 419$) in
Fig.~\ref{pattern-fig}b.  Thus, if one can find a proper $\lambda$
that falls in this range, then one can claim that the estimation of
$\Delta$ is ``robust''. But the range of this pattern may change for
other $M$ and $k$ parameters, in which case there is no guarantee that
the estimated $\lambda$ falls in this range. Furthermore, 
no matter which method
is used for assessing the goodness of fit, 
if we test $\Delta$ in a specific range
with a fixed $\lambda$, then we may come up with different 
values of $\theta$ --
some inside the pattern, some outside, none inside, etc.

Instead of $\lambda$, we use $\theta$ as a regularisation parameter in
\S \ref{evo-section}. In fact, we aim to create an automatic algorithm
that performs a global search for all parameters, and then finds the
proper $\theta$ that falls in the pattern; with our EA, we have done
this (see \S \ref{evo-section}).  A review of other general
regularisation techniques for inverse problems can be found in Conan
\cite{Cowan:1998:SDA}.

\subsection{Optimisation}\label{optimisation-section}
The above formulation can be seen as an optimisation problem where the variables are $\Delta$, $k$ and $\theta$. Conventional gradient-based optimisation techniques cannot be used since the above kernel-based approach is not differentiable with respect to the discrete variables $k$ and $\theta$, regardless of the loss function.

Of course, since both $k$ and $\theta$ are finite range discrete variables,
one can employ a brute force search driven by cross-validation.
Apart form having to deal in a systematic and time-consuming manner 
with a huge search space,
it is also not clear what the appropriate ranges and resolutions
should be for parameters such as $\Delta$ or $k$.
In any case, we compare our EA approach (see section \S\ref{evo-section})  with 
a non-evolutionary kernel-based approach (K-V) for estimating  time delay $\Delta$ by cross-validating $k$ and the regularisation parameter $\lambda$ \cite{Cuevas:2006:HAT,Cuevas:2006:ECML}.

\begin{figure}
\centering
\includegraphics[width=3.5in]{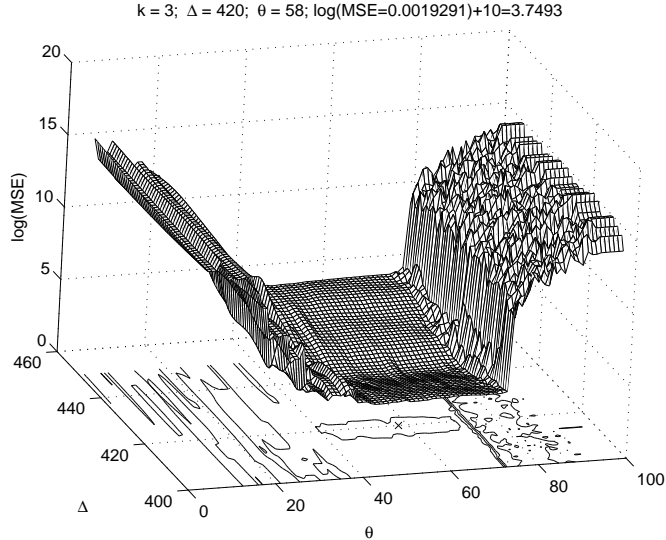}
\caption{Example of Search Landscape. The data is from the doubly-imaged quasar Q0957+561, in the $g$-band. The parameters $k$ was fixed to 3, and the offset $M$ to 0.117. Algorithm \ref{fitness} is used to obtain the log(MSE); see \S \ref{evo-section}. The surface has been shifted upwards by 10 units for visualisation.}
\label{landscape-fig}
\end{figure}

An example of the search landscape is in Fig \ref{landscape-fig}. We use $\Delta_{min}=400$ and $\Delta_{max}=450$ with unitary increments, and $\theta=1,2,...,n$, where $n=97$. The parameter $k$ is fixed to 3, and the offset $M$ to 0.117. The real optical data ($g$-band) in \S \ref{data-section} is used. Then, Algorithm \ref{fitness} (explained later in \S \ref{evo-section}) is applied to obtain the MSE$_{CV}$. Note that this landscape may change for other $k$ values and for other data sets. We can see that from $\theta=80$ to $n$ the error surface is quite complicated for simple search algorithms, e.g., gradient descent (if differentiable), hill climbing or simulated annealing search. There are also more local minima when $\theta < 45$; see also Fig. \ref{pattern-fig}. In the $\theta$-$\Delta$ plane, the mark (x) shows the best parameter combination; i.e., minimum MSE$_{CV}$. To smooth the surface and help the visualisation, we use a logarithmic scale. 

\section{Evolutionary Algorithm (EA)}\label{evo-section}

Following the kernel-based approach in \S\ref{kernel-section}, we
have three parameters: (i) the time delay $\Delta$; (ii) the variable
width $k$; and (iii) the number of singular values to retain
$\theta$. Besides, we have a measurement of fitness (objective function),
e.g. log-likelihood or a loss function, which, along with the others,
gives us a third-dimensional search space $\Phi$. Therefore, we follow an EA to avoid local
minima \cite{Biethahn:1995:EA,Goldberg:1989:GA,Spears:2000:EA}.

Let us define as our population
\begin{equation}
  \label{P-equation}
   \vec{P_\ell} = \left [ \begin{array}{lll | l}
                                \Delta_1 &  \theta_1 & k_1 & f_1 \\
                                \Delta_2 &  \theta_2 & k_2 & f_2 \\
                                ...      & ... & ...      & ... \\
                                \Delta_x &  \theta_x & k_x &  f_x \\
                                ...      &  ... & ...      & ... \\
                                \Delta_p &  \theta_p & k_p &  f_p \\
                          \end{array}
                  \right ],
\end{equation}
\noindent 
where each row in $\vec{P_\ell}$ is a hypothesis commonly referred to as
individual or chromosome, which is a set of parameters \{$\Delta_x,
\theta_x, k_x$\}, randomly initialised. 

Then we have $p$ hypotheses \cite{Mitchell:1997:book}. Each hypothesis
$x$ is evaluated by $f_x$, which is a measure of fitness. For this, we
use the mean squared error (MSE$_{CV}$) given by Cross Validation
(CV), in Algorithm \ref{fitness}, where $T=O-V$ is the training set,
$O$ is the set of all observations such as
$O=\{(x_A(t_i),x_B(t_i))|i\}$, and $V$ is the validation set (the
log-likelihood or simple mean squared error can also be used, but this
might lead to overfitting).  Thereafter, we apply artificial genetic
operators such as selection, crossover, mutation and reinsertion to
generate $\vec{P_2},...,\vec{P_g}$ populations. At the $g$ generation,
we choose from $\vec{P_g}$ the best set of parameters (or individual)
according to its fitness; i.e., with minimum $f_x$.  This procedure is
summarised in Algorithm \ref{pseudocode-ea}, and the details are in
\S\ref{rep-evo-section}.

\begin{algorithm}
\caption{Fitness function ($\Delta_x$, $k_x$, $\theta_x$)}
\label{fitness}
 $Blocks \leftarrow 5$ \\
 $PointsPerBlock \leftarrow n/Blocks$ \\
 {\bf for} $i \leftarrow 1$  \hbox{\rm \bf to} $PointsPerBlock$ \\
      \{\\
      \hspace*{0.5in} Remove the $i^{th}$ observation of each block and include it \\
      \hspace*{0.5in} in the validation set $V$ \\
      \hspace*{0.5in} Compute $\vec{h_{A}}$ and $\vec{h_{B}}$ for the training set $T$ \\
      \hspace*{0.5in} Get MSE$_{CV}$ on the validation set $V$  \\
      \hspace*{0.5in} $R(i) \leftarrow$ MSE$_{CV}$ \\
   \} \\
 $f_x \leftarrow mean(R)$  \\
  return $f_x$ \\
\end{algorithm}

This process leads to artificial evolution, which is a stochastic
global search and optimisation method based on the principles of
biological evolution \cite{Goldberg:1989:GA,Spears:2000:EA}.

\begin{algorithm}
\caption{Evolutionary Algorithm}
\label{pseudocode-ea}
 {\it (See \ref{rep-evo-section} for details)}  \\
 Initialise population $\vec{P_1} = [\vec{P_1^1} \ \vec{P_1^2}]$ \\
 Evaluate population  $\vec{P_1}$ with Algorithm \ref{fitness}\\
 {\bf for} $\ell \leftarrow 2 \ \hbox{\rm \bf to} \ g$ \\
    \{ \\
      \hspace*{0.5in} Select  $\vec{P_\ell^{1'}}$ and $\vec{P_\ell^{2'}}$ from $\vec{P_1}$ \\
      \hspace*{0.5in} Recombine $\vec{P_\ell^{1'}}$,  Recombine $\vec{P_\ell^{2'}}$  \\
      \hspace*{0.5in} Mutate  $\vec{P_\ell^{1'}}$ \,  Mutate  $\vec{P_\ell^{2'}}$ \\
      \hspace*{0.5in} Evaluate $\vec{P_\ell^{'}} = [\vec{P_\ell^{1'}} \ \vec{P_\ell^{2'}}]$ with Algorithm \ref{fitness}\\
      \hspace*{0.5in} Reinsert $\vec{P_\ell^{'}}$ into $\vec{P_{\ell-1}}$ to obtain $\vec{P_\ell}$ \\
   \}
\end{algorithm}

\subsection{Representation and Evolution Operators}\label{rep-evo-section}
We represent every population $\vec{P_\ell}$ in every generation  $\ell = 1,2,...,g$,
as two linked 
populations of the same size $p$, $\vec{P_\ell} = [\vec{P_\ell^1} \ \vec{P_\ell^2}]$. 
The $x$-th individual in population 
$\vec{P_\ell}$ corresponds to 
the $x$-th individual in populations $\vec{P_\ell^1}$ and
 $\vec{P_\ell^2}$.
The population $\vec{P^1_\ell}$ uses reals to represent
$\Delta_x$, while  $\vec{P_\ell^2}$ employs integers to represent $\theta_x$
and $k_x$. First we initialise randomly $\vec{P_1}$ and evaluate the population with the above fitness function.
Second, 
we select half of the population $\vec{P_\ell}$ for reproduction. 
This selection of individuals is then applied 
to each sub-population $\vec{P_\ell^1}$, $\vec{P_\ell^2}$,
i.e. the indexes of selected individuals
in both sub-populations are the same.
We use roulette wheel selection to obtain $\vec{P_\ell^{1'}}$ and $\vec{P_\ell^{2'}}$. 
Third, we apply individually recombination and mutation on 
$\vec{P_\ell^{1'}}$ and $\vec{P_\ell^{2'}}$.
Finally, we evaluate the new linked population
$\vec{P_\ell^{'}} = [\vec{P_\ell^{1'}} \ \vec{P_\ell^{2'}}]$ to obtain its fitness and perform reinsertion of offsprings between $\vec{P_\ell}$ and $\vec{P_\ell^{'}}$ (elitist strategy).
We repeat the above procedure until $\vec{P_g}$ is obtained. Note that $\vec{P_1}$ to $\vec{P_g}$ have always the same size.

We use linear recombination and $mutbga$ as mutation\footnote{We also
tested Gaussian mutation, which leads to a similar performance.} (as in
Breeder Genetic Algorithm \cite{GA:1996:toolbox}) for $\vec{P_1^{1'}}$,
and double-point crossover and discrete mutation for $\vec{P_1^{2'}}$. In
both cases, we use 0.5 as mutation rate. We employ a population size of
$p=300$ individuals and $g=50$ generations, unless other values are
given. The above evolutionary algorithm\footnote{We use the Genetic
Algorithm Toolbox for MATLAB
\cite{GA:1996:toolbox,Chipperfield:1994:GA}, which is available online
with good documentation.} is what we refer hereafter as EA (with
mixed representation unless otherwise stated).

Moreover, we evolved the $M$ parameter, and, rather than mixed types, we
also tested only real representation with a single population performing two kinds of flooring
for integers, in the population and in the fitness function.

\section{Results}\label{results-section}
Here we present the results of the application of our evolved kernel
approach on real and artificial data. 
We compare the performance of our EA, on the same data sets, against
two of the most popular methods from the astrophysics literature: (a)
{\it Dispersion spectra} method
\cite{Pelt:1994:TDC,Pelt:1996:TLC,Pelt:1998:EMT}, and (b) the
structure-function-based method (PRH, \cite{Press:1992:TTD}).  In addition, we compare with 
the performance of our previous approach, based on 
kernels with variable width (K-V)
\cite{Cuevas:2006:HAT}, which was applied to a
different (radio) observational data set and one of the synthetic 
data sets (large scale data only).

Two versions of Dispersion spectra are used;
$D_1^2$ is free of parameters
\cite{Pelt:1994:TDC,Pelt:1996:TLC} and $D_{4,2}^2$ has a decorrelation
length parameter $\delta$ involving only nearby points in the weighted
correlation \cite{Pelt:1996:TLC,Pelt:1998:EMT}. For the case of the
PRH method, we use the image A from the data to estimate the structure
function \cite{Press:1992:TTD}. In the last subsection, we compare EA
against a Bayesian method on data sets generated by this Bayesian
approach \cite{Harva:2006:IEEE}.

In the first section below, we present the results of our analysis
of the real observational data, followed by the results from the various
synthetic data: large scale, PRH and Wiener data (see
\S\ref{data-section})

\subsection{Astronomical observations}
Here we use the observational optical data outlined in
\S\ref{optical-section}. We begin by showing the results of our EA
evolving $M$ with real representation only. The integer parameters 
are floored at fitness function. For $M$, and each parameter
in Eq.~(\ref{P-equation}), we define the following general bounds:
$\Delta=[400, 450]$, $k=[1, 15]$, $\theta=[1, n]$, and $M=[0.10,
0.20]$. The results of ten runs (realisations) are given in
Table~\ref{result1-table}. The set \{$\Delta$, $M$, $\theta$, $k$\} is
the best solution (individual) at $g=50$ according to $f_x$ (i.e.,
MSE$_{CV}$). The column {\it Convergence} 
shows at what generation a
stability has been reached, i.e., from what generation the
MSE$_{CV}$ is constant.

\begin{table}
\renewcommand{\arraystretch}{1.3}
\caption{Evolutionary algorithm with all reals: results on DS1}
\label{result1-table}
  \begin{tabular}{ c c c c c c c } \hline
Run & $\Delta$ & $M$ & $\theta$ & $k$ & $f_x$ & Convergence at \\ \hline

1       &419.67 &0.1495 &58     &3      &0.0019249601   &40 \\
2       &419.67 &0.1462 &58     &3      &0.0019249602   &43 \\
3       &419.67 &0.1923 &58     &3      &0.0019249605   &40 \\
4       &419.68 &0.1398 &58     &3      &0.0019249620   &46 \\
5       &419.68 &0.1217 &58     &3      &0.0019249577   &38 \\
6       &419.68 &0.1197 &58     &3      &0.0019249593   &33 \\
7       &419.68 &0.1733 &58     &3      &0.0019249592   &28 \\
8       &419.68 &0.1516 &58     &3      &0.0019249615   &37 \\
9       &419.67 &0.1482 &58     &3      &0.0019249588   &47 \\
10      &419.68 &0.1656 &58     &3      &0.0019249586   &40 \\ 
      \hline 
    \multicolumn{7}{c}{$\Delta$ is given in days}
  \end{tabular}
\end{table}

Of the continuous optimisation approaches, we tested one, (1+1)ES
\cite{Rowe:2004:ES}, which is based on the Gray-code neighbourhood
distribution, and uses real representation. 
(1+1) means that one parent is selected and one child is produced in each single step of the Evolutionary Strategy (ES).
We chose the (1+1)ES approach
because
our fitness function is costly, so one expects to require fewer fitness evaluations
than our EA. Rowe et al. \cite{Rowe:2004:ES} have shown superior performance of their (1+1)ES
over Improved Fast Evolutionary Programming (IFEP), on some benchmark
problems, and on a real-world problem (medical tissue optics). IFEP is
also a continuous optimisation approach \cite{Yao:1999:IFEP}. For
(1+1)ES, the precision is set to 200, and variable bounds set as
above, allowing until 15,000 iterations. The convergence is reached
after 14,410 iterations by using the same fitness function (Algorithm
\ref{fitness} in \S\ref{evo-section}), so we also floor at fitness
function for integer variables. This ES yields $\Delta=419.6$,
$M=0.1732$, $\theta=58$, $k=3$, and MSE$_{CV}=1.9249617 \times
10^{-3}$.

In Table~\ref{result2-table}, we present ten runs, resulting from our
EA with mixed types as discussed in \S\ref{rep-evo-section}. Here, $M$
is not evolved, being fixed to 0.117. The variable bounds are also set
as above. Table~\ref{result1-table} shows that regardless of the value
of $M$, the time delay $\Delta$ is consistent, this justifies that $M$
does not need to be evolved. Rather, we use the reported value
$M=0.117$ \cite{Kundic:1997:ARD}.

\begin{table}
\renewcommand{\arraystretch}{1.3}
\caption{Evolutionary algorithm with mixed types: results on DS1}
\label{result2-table}
  \begin{tabular}{ c c c c c c } \hline
Run & $\Delta$ &  $\theta$ & $k$ & $f_x$ & Convergence at \\ \hline

1       &419.68 &58     &3      &0.0019249744   &42 \\
2       &419.67 &58     &3      &0.0019249722   &34 \\
3       &419.69 &58     &3      &0.0019249722   &47 \\
4       &419.67 &58     &3      &0.0019249719   &49 \\
5       &419.66 &58     &3      &0.0019249691   &40 \\
6       &419.66 &58     &3        &0.0019249670   &45 \\
7       &419.66 &58     &3      &0.0019249753   &44 \\
8       &419.67 &58     &3      &0.0019249724   &47 \\
9       &419.47 &71     &3      &0.0018908716   &32 \\
10      &419.67 &58     &3      &0.0019249711   &49 \\ 
      \hline 
    \multicolumn{6}{c}{$\Delta$ is given in days}
  \end{tabular}
\end{table}

We point out that in Tables~\ref{result1-table} and
\ref{result2-table} the EA suggests $\theta=58$, which falls within the
pattern in Fig.~\ref{pattern-fig}b.

In Table~\ref{result1-table}, we can see that the parameter $M$ is not
crucial in the time delay estimation. Therefore, in
Table~\ref{result2-table}, we omit it. The results of these tables
yield similar $\Delta$ estimates regardless of the representation
(reals or mixed types).

The results of the (1+1)-ES are also consistent, even though (1+1)-ES
requires a larger number of iterations. On the one hand, for our EA,
when $g=50$ (maximum number of generations), we perform 7,800
evaluations of the fitness function, because of our elitist
strategy. On the other hand, (1+1)-ES tends to converge in around 14,000
iterations across different initialisations. Every iteration corresponds
to a fitness evaluation. (1+1)-ES demands more
computational time (about twice as much) and therefore we do not use this algorithm to
analyse artificial data. Since we use the same fitness function, one
would expect to get similar performance to the EA. Moreover, a theoretical analysis 
in multi-objective optimisation suggests a better performance of  population-based 
algorithms (such as the EA used here), compared with (1+1)ES \cite{Giel:2006:OEP}.

In the astrophysics literature,
the best (smallest quoted error)
previous measures for this time delay can be found to be  417$\pm$3
days \cite{Kundic:1997:ARD} and 419.5$\pm$0.8 days
\cite{Cuevas:2006:ECML}. Therefore, the results in Tables
\ref{result1-table} and \ref{result2-table} are consistent. However,
we think that the estimate of 417$\pm$3 days, from this data set, is
underestimated because, for the quasar Q0957+561, the latest reports
also give estimates around 420 days by using other data sets
\cite{Ovaldsen:2003:NAP}. One is reminded that the gravitational lensing
theory predicts that the time delay must be the same regardless of the
wavelength of observation
\cite{Refsdal:1966:R66,Kochanek:2004:THC,Saha:2000:GL}.

\subsection{Artificial Data}
For the analysis of real data presented above, we do not know the
actual value of the time delay. In order to evaluate the relative
performance of various methods, we therefore present the analysis of 
synthetic data sets, produced from a set of known parameters.

\subsubsection{Large Scale Data}\label{large-scale-data-results-section}
In all cases, the time delay under analysis is given by 
trials of values 
between $\Delta_{min}=0$ and $\Delta_{max}=10$ (also bounds in
our EA), with increments of 0.1, where the ratio $M$ is set to its true
value $0.1$. The parameter $\delta$ is set to 5, for
$D_{4,2}^2$. When using the PRH method, we use bins in the range of [0, 10]
for estimating the structure function from the light curve
of Image~A. In our EA, besides the above
$\Delta$ bounds, we use the following bounds: $\theta=[1, n]$, and
$k=[1, 15]$. For K-V, we cross-validate $k$ and $\lambda$; the ranges
are $k=[1, 15]$ and $\lambda=[10^{-1},10^{-2},...,10^{-6}]$
(see \S\ref{regularisation-section}).

Table~\ref{result3-table} presents the results for all time delay
estimates; i.e., $\eta = 38,505$. The best results are highlighted
in bold. Regarding the statistics in Table~\ref{result3-table},
let $\hat{\Delta}_j$, $j=1,2,...,\eta$, be estimated time delays, where $\eta$ is the quantity of time delay estimates.
The empirical mean is 
\begin{equation}
\label{mean-estimates}
     \hat{\mu} = \frac{1}{\eta}\sum_{j=1}^{\eta} \hat{\Delta}_j,
\end{equation}
\noindent 
and the empirical standard deviation is 
\begin{equation} 
\label{std-estimates}
     \hat{\sigma} = \sqrt{\frac{1}{\eta-1}\sum_{j=1}^{\eta} (\hat{\Delta}_j - \hat{\mu})^2}.
\end{equation}
The estimators $\hat{\mu}$ and $\hat{\sigma}$ are used to estimate the bias and variance of time delay estimates, respectively. The mean squared error is given by
\begin{equation} 
\label{mse-estimates}
     {\rm MSE} = \frac{1}{\eta} \sum_{j=1}^{\eta} (\hat{\Delta}_j - \mu_0)^2,
\end{equation}
\noindent where $\mu_0$ is the true time delay. The average of absolute error is
\begin{equation}
\label{ae-estimates}
     {\rm AE} = \frac{1}{\eta} \sum_{j=1}^{\eta} |\hat{\Delta}_j - \mu_0|.
\end{equation}
The 95\% confidence intervals (CI) for $\hat{\mu}$ are given by $\hat{\mu}
\pm 1.96 \times \hat{\sigma} / \sqrt{\eta}$, where the constant
depends on the desired confidence level and the sample size; e.g., see
Table~IIIa in \cite{Bery:1990:STM}.

\begin{table}
\renewcommand{\arraystretch}{1.3}
\caption{Large scale data results: statistical analysis}
\label{result3-table}
 \begin{tabular}{l r r r r r}
 \hline
  Statistic      & $D_1^2$    & $D_{4,2}^2$  & PRH        & K-V        &  EA        \\ \hline
   95\% CI     &{\bf [5.00,}& [5.58,       & [2.67,     & [4.94,     &{\bf [5.00,}   \\
               & \ {\bf 5.02]}& \ 5.59]    & \ 2.73]    & \ 4.95]    & \ {\bf 5.02]} \\
    CI range     & 0.02       &{\bf 0.01}    & 0.06       &{\bf 0.01}  & 0.02          \\ 
     MSE          & 0.74       & 0.99         & 13.46      &{\bf 0.47}  & 0.63       \\
     AE          & 0.52       & 0.59         & 3.01       &{\bf 0.39}  & 0.41        \\ 
  $\hat{\mu}$    &{\bf 5.013} & 5.589        & 2.704      & 4.946      & 5.015       \\
  $\hat{\sigma}$ & 0.86       & 0.80         & 2.86       &{\bf 0.68}  & 0.79        \\
\hline \hline
\end{tabular}
\end{table}

\begin{figure*}[ht!]
\centering
\includegraphics[width=5.5in]{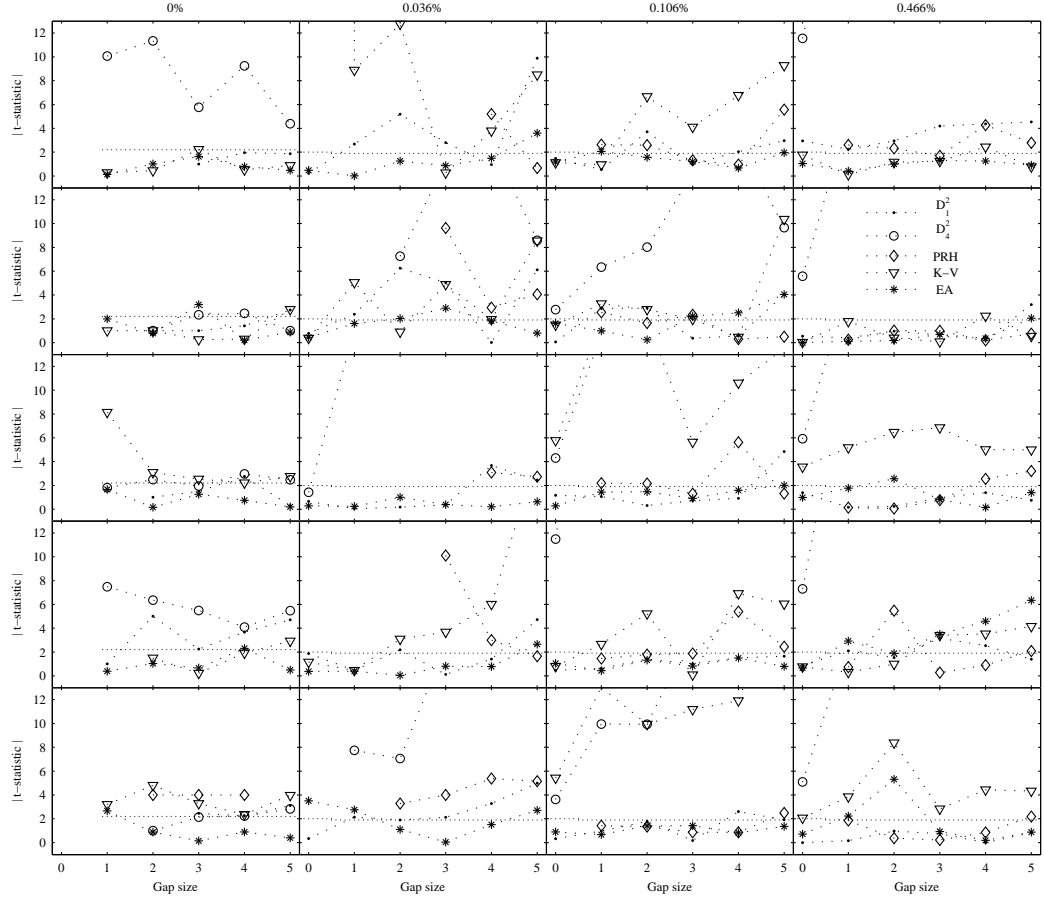}
\caption{The $t$-test results applied to artificial Large Scale data. Each row corresponds to a different underlying
function (DS-5-1,DS-5-2,...,DS-5-5), and each column corresponds to a different level of noise (0\%, 0.036\%, 0.106\% and 0.466\%). Every plot shows the results of $|{\cal T}|$ from five methods; i.e., $D_1^2$, $D_{4,2}^2$, PRH, K-V and EA; shaded point, circle, diamond, triangle and asterisk respectively. Note that all the plots have the same scale on the  $y$-axis. See \S\ref{large-scale-data-results-section} for details.}
\label{t-test-fig}
\end{figure*}

We also performed a $t$-test on these results, where the hypothesis to
test is $H_0$:~$\mu_0=5$. The results are shown in
Fig.~\ref{t-test-fig}, where the estimates are grouped by the
underlying function, the level of noise and gap size. Since ${\cal T}$
follows a Student's $t$-distribution,
which is centred at zero, those
values close to zero are statistically significant
\cite{Anderson:1978:ISAD,Dudewicz:1988:MMS,Bery:1990:STM}. The
horizontal dotted line shows the threshold for a significance level of
95\%, $\alpha=0.05$; i.e., when ${\cal P} < \alpha$, where ${\cal P}$ is known as the p-value. Thus, the
threshold values for $|{\cal T}|$ in Fig.~\ref{t-test-fig} are 2.2, 2
and 1.9 for $\nu=\{9,49,499\}$, degrees of freedom, respectively (see
Table~\ref{large-data-table}).

In Table~\ref{result3-1-table}, we show the number of cases that
satisfy the above threshold values. In other words, see Fig.~\ref{t-test-fig} and count the number of points below the horizontal dotted line per level of noise-- the significant results. In Table~\ref{result3-1-table}, the results are grouped by noise level only, and the best ones are highlighted in bold. We also tested the significance of time delay estimates with nonparametric hypothesis testing, such as sign test and Wilcoxon's signed-rank test \cite{Bery:1990:STM}, with similar results.

\begin{table}
\renewcommand{\arraystretch}{1.3}
\caption{Large scale data results: $t$-test}
\label{result3-1-table}
 \begin{tabular}{c c c c c}
 \hline 
  Method       & 0\%  & 0.036\% & 0.106\% & 0.466\% \\ \hline
   $D_1^2$     &   10 &   13    &   21    &   20    \\
   $D_{4,2}^2$ &   6  &   1     &   0     &   0     \\
   PRH         &   0  &   2     &   14    &   16    \\
   K-V         &   11 &   5     &   6     &   13    \\
   EA          & {\bf 22} & {\bf  23 }    &  {\bf 24 }    & {\bf  22} \\
\hline 
\multicolumn{5}{l}{see \S\ref{large-scale-data-results-section} for details}
\end{tabular}
\end{table}

\begin{figure*}[ht!]
\centering
\includegraphics[width=4.5in]{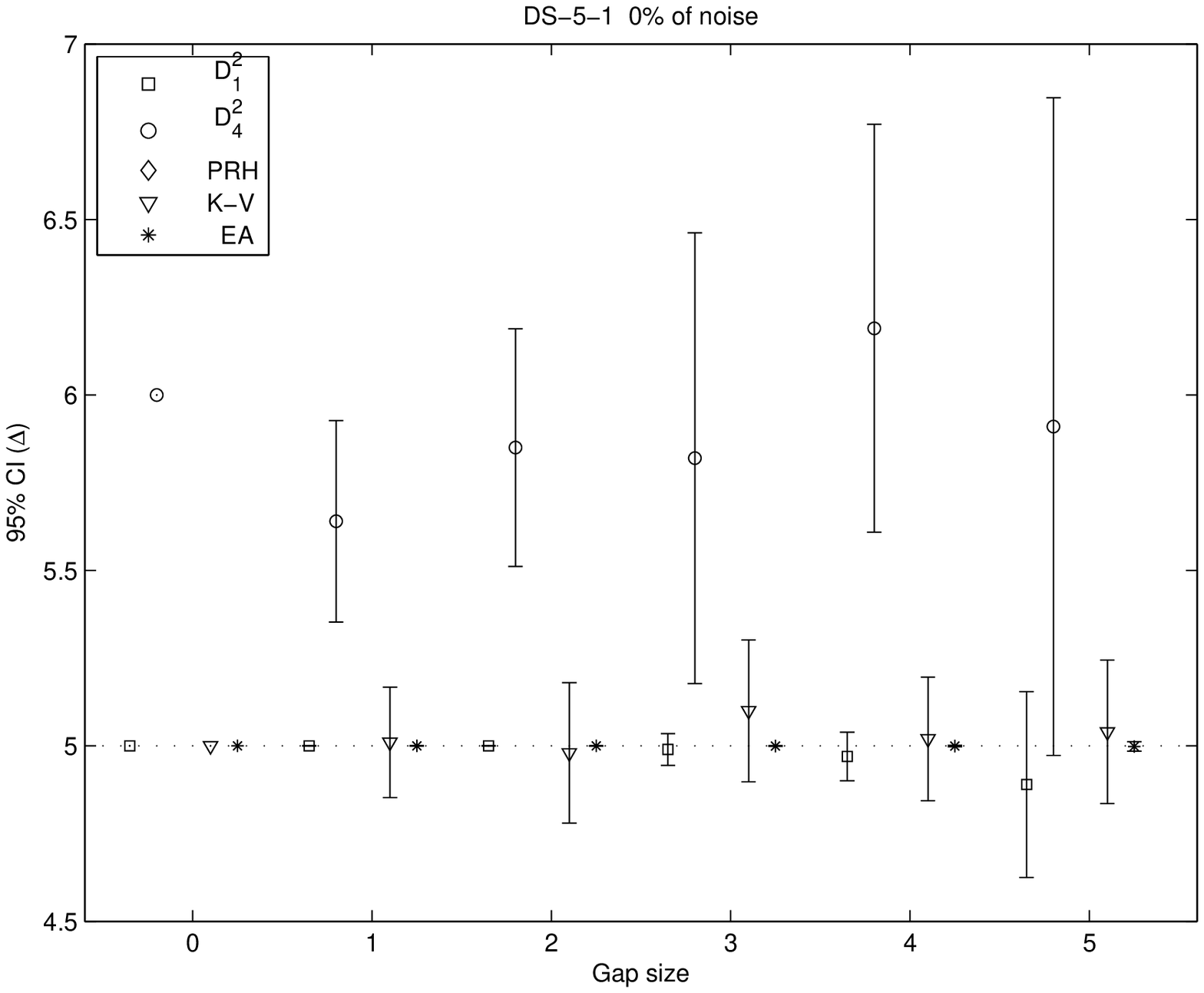}
\caption{95\% Confidence Intervals on Large Scale Data, DS-5-1 with 0\% of noise only. This plot shows the 95\% CI from five methods: $D_1^2$, $D_{4,2}^2$, PRH, K-V and EA. The PRH intervals are not visible here because are below the bound of 4.5 days (see Table \ref{result3-table}). We use this bound for visualisation purposes. The horizontal and dotted line is located at the true delay, $\Delta=5$.}
\label{CI-fig}
\end{figure*}

Like Table~\ref{result3-1-table}, Table~\ref{result3-2-table} shows the quantity of cases where the true delay, $\Delta=5$, falls within the 95\% CI. The results are also grouped by noise level. In Fig.~\ref{CI-fig}, we illustrate the 95\% CI for DS-5-1, i.e., one underlying function with 0\% of noise only.

\begin{table}
\renewcommand{\arraystretch}{1.3}
\caption{Large scale data results: 95\% CI}
\label{result3-2-table}
 \begin{tabular}{c c c c c}
 \hline 
  Method       & 0\%  & 0.036\% & 0.106\% & 0.466\% \\ \hline
   $D_1^2$     & 23       &   14    &   22     &   20      \\
   $D_{4,2}^2$ & 12       &   4     &   0      &   0       \\
   PRH         & 0        &   0     &   0      &   6       \\
   K-V         & 19       &   6     &   6      &   13      \\
   EA     & {\bf 27} &{\bf  23}& {\bf 25 }& {\bf  22} \\
\hline 
\multicolumn{5}{l}{see \S\ref{large-scale-data-results-section} for details}
\end{tabular}
\end{table}

\begin{figure*}[ht!]
\centering
\includegraphics[width=5.5in]{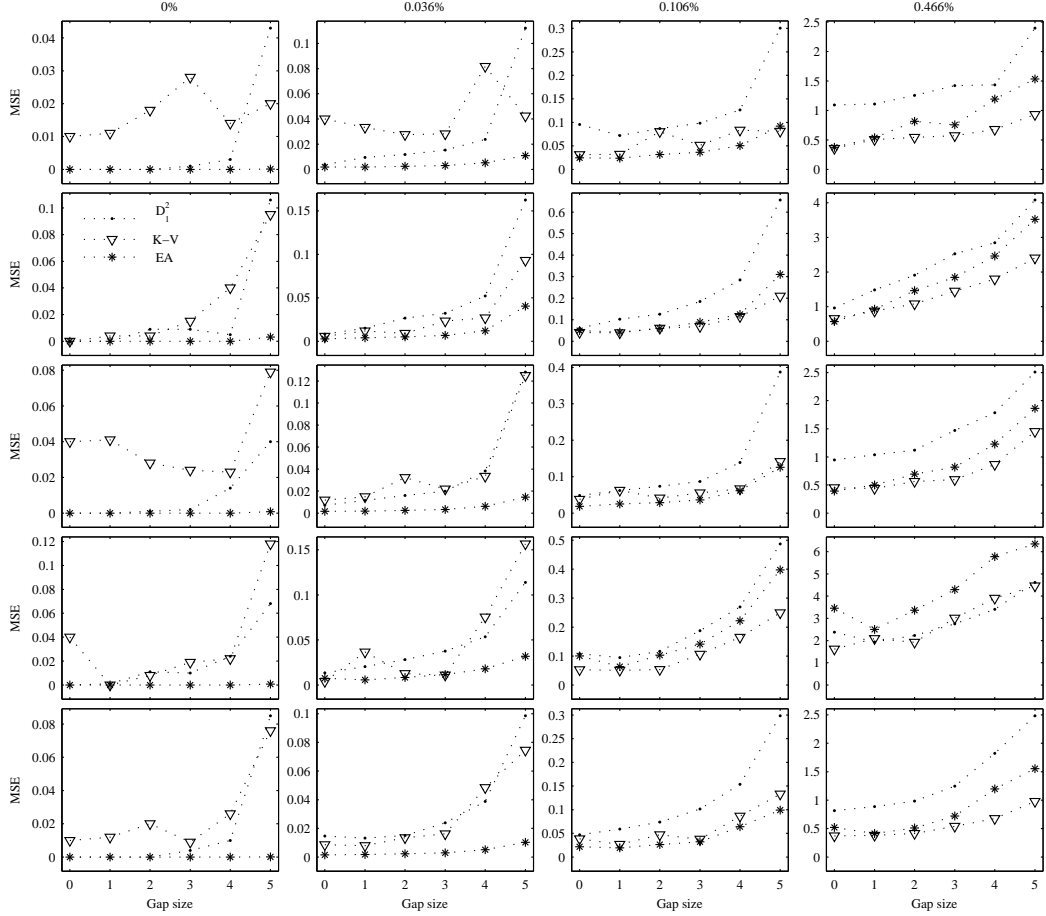}
\caption{The use of MSE on artificial Large Scale Data. Each row corresponds to a different underlying
function (DS-5-1,DS-5-2,...,DS-5-5), and each column corresponds to a different level of noise (0\%, 0.036\%, 0.106\% and 0.466\%). Every plot shows the results of MSE statistic from three methods; i.e., $D_1^2$, K-V and EA, indicated by shaded point, triangle and asterisk, respectively.}
\label{MSE-fig}
\end{figure*}

In Fig.~\ref{MSE-fig} are shown the results of MSE
(\ref{mse-estimates}), where the estimates are grouped by level of noise as in the previous figure --
Fig.~\ref{t-test-fig}. Here, for low levels of noise: 0\% and 0.036\%, the asterisk (the proposed EA) has and outstanding performance because the MSE is close to zero. The AE statistic (\ref{ae-estimates}) gives
similar results with this grouping (not shown).

From Table~\ref{result3-table}, the best results are for $D_1^2$, K-V
and EA. Since the noise is about 0.01 mag ($<0.106\%$) (standard
deviation) in the observational optical data, we are interested in
exploring the effects of various levels of noise. Therefore, in
Table~\ref{result3-3-table}, we show the results of the estimates
grouped by noise level, regardless of the gap size. 
The best results are highlighted in bold. As in Tables \ref{result3-1-table} and \ref{result3-2-table}, and Figure \ref{MSE-fig}, the results from EA are promising.

\begin{table}[!ht]
\renewcommand{\arraystretch}{1.3}
\caption{Large scale data results grouped by noise level}
\label{result3-3-table}
 \begin{tabular}{l r r r r}
 \hline 
                   &  \multicolumn{4}{c}{Noise Level} \\
  Statistic        &  0\%     & 0.036\%   &  0.106\%   & 0.466\%  \\ \hline
  \multicolumn{5}{l}{ {\bf Method:} {\boldmath $D_1^2$} } \\
    MSE            & 0.017    & 0.044     & 0.182     & 2.014    \\
     AE            & 0.060    & 0.147     & 0.321     & 1.121    \\ 
  $\hat{\mu}$      & 4.95     & 4.98      & 4.98      & 5.07     \\
  $\hat{\sigma}$   & 0.12     & 0.20      & 0.42      & 1.41     \\ \hline

  \multicolumn{5}{l}{ {\bf Method: K-V} } \\
    MSE            & 0.029    & 0.041     &{\bf 0.084} &{\bf 1.312}  \\
     AE            & 0.117    & 0.139     & 0.219      &{\bf 0.833}  \\ 
  $\hat{\mu}$      & 4.93     & 4.94      & 4.93       &{\bf 4.96}   \\
  $\hat{\sigma}$   & 0.11     & 0.13      &{\bf 0.21}  &{\bf 0.83}   \\ \hline

  \multicolumn{5}{l}{ {\bf Method: EA} } \\
    MSE            &{\bf 1.9e-4}&{\bf 0.008} & 0.090     & 1.831    \\
     AE            &{\bf 4.7e-3}&{\bf 0.066} & {\bf 0.216}& 0.984    \\ 
  $\hat{\mu}$      &{\bf 4.99}  &{\bf 4.99}  & {\bf 4.99} & 5.05     \\
  $\hat{\sigma}$   &{\bf 0.01}  &{\bf 0.09}  & 0.30      & 1.35     \\ \hline

   $\eta$          &  255     &  12,750   & 12,750      & 12,750  \\ 

\hline 
\end{tabular}
\end{table}

\begin{figure*}[ht!]
\centering
\includegraphics[width=5.5in]{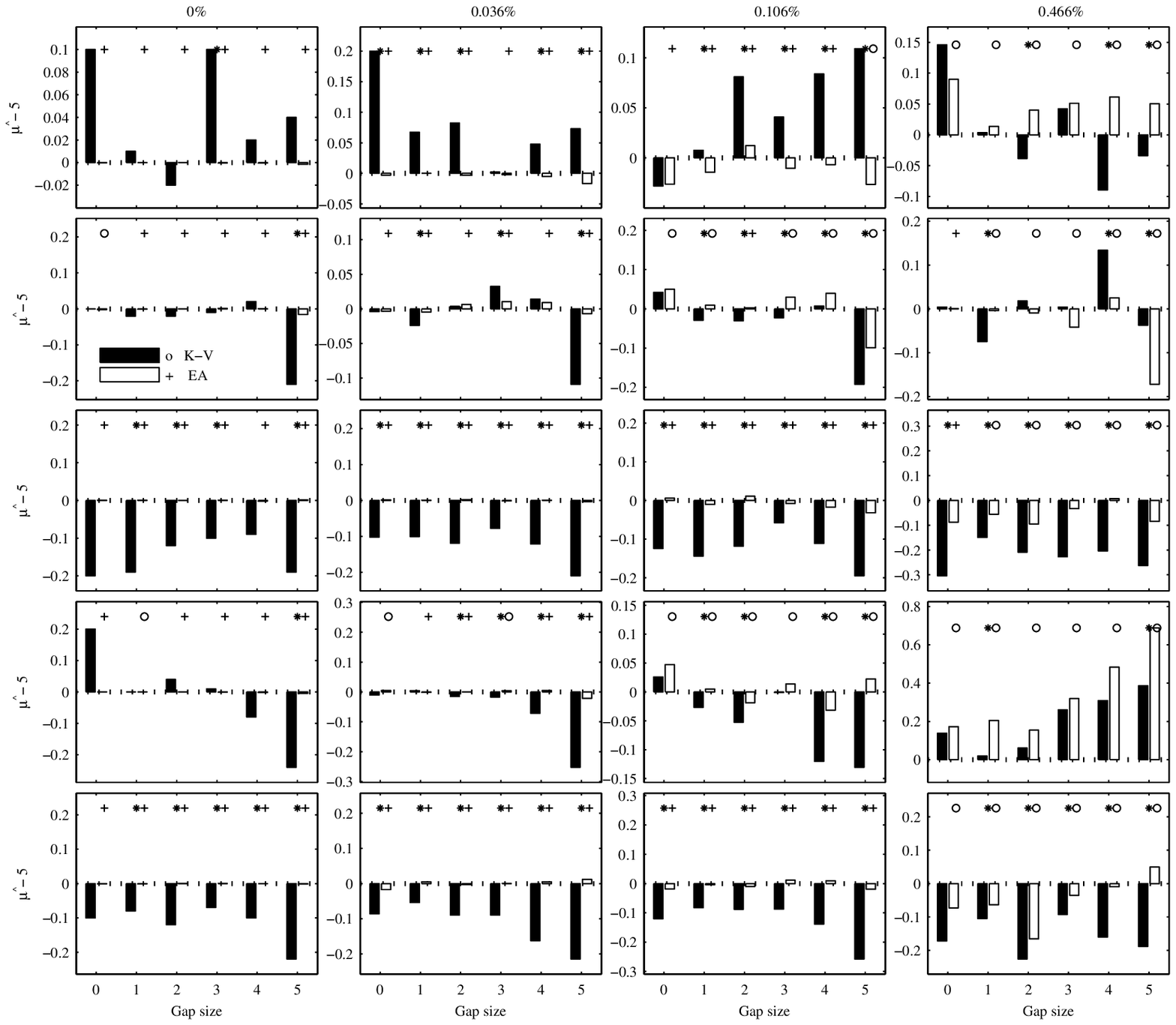}
\caption{Paired $t$-test on Large Scale Data. The paired $t$-test is performed on time delay estimates from K-V and EA. Bars represent the estimator $\hat{\mu}-\mu_0$. At the top of each plot, a circle appears when ${\rm MSE}_{K-V}< {\rm MSE}_{EA}$, and the plus symbol when ${\rm MSE}_{K-V} > {\rm MSE}_{EA}$. The asterisk at top means that the p-value resulting from paired $t$-test is less than 0.05, i.e., 95\% confidence level.}
\label{p-t-test-fig}
\end{figure*}

Finally, we compare the performances of methods $D_1^2$, K-V and EA through paired tests on time delay estimates and MSE. As an example, in Fig. \ref{p-t-test-fig}, we show K-V against EA with paired $t$-test. The bars represent the mean estimator $\hat{\mu}-\mu_0$ for each method, where $\hat{\mu}$ is the mean of time delay estimates, and $\mu_0$ is the true time delay. At the top of each plot appears either a circle or a plus symbol representing K-V and EA, respectively. If ${\rm MSE}_{K-V}< {\rm MSE}_{EA}$ then a circle appears, and we display the plus symbol when ${\rm MSE}_{K-V} > {\rm MSE}_{EA}$. The asterisk at the top means that the difference is significant at the level ${\cal P}< 0.05$, i.e., 95\% confidence level. In simple words, if the bar is large then the results are bad, because one is far from the true value $\mu_0=5$. Note that when the noise is low, 0\%, 0.036\% and 0.106\%, the empty bars (EA) are small, that is, when the plus symbol (+) appears at the top. Therefore the results from EA are interesting. Moreover, note that the asterisk (*) at the top appears in some cases. The asterisk means that the comparison is significant when performing the paired $t$-test with a 95\% confidence level.

In Table \ref{result3-4-table}, we summarise the number of cases where ${\cal P}< 0.05$, including paired sign test and paired-sample Wilcoxon signed-rank test. We also compare the results from $D_1^2$ with K-V and EA. For each comparison, the pairs are represented by $M_1$ (o) and  $M_2$ ($+$). The columns corresponding to $t$-test show two quantities. The first one corresponds to the quantity of symbols ``o'' ($M_1$) with an asterisk also. The second one is the quantity of  symbols ``+'' ($M_2$) with asterisks. To obtain these numbers one should imagine a figure similar to Fig. \ref{p-t-test-fig} for the methods involved. The columns for sign test and signed-rank test are obtained in a similar manner; rather than $t$-test in Fig. \ref{p-t-test-fig}, were used other tests. Note again that the best results are from EA.

\begin{table}
\renewcommand{\arraystretch}{1.3}
\caption{Large scale data results: paired tests}
\label{result3-4-table}
 \begin{tabular}{c c c c c c c c}
 \hline 
 &  & \multicolumn{2}{c}{$t$-test} & \multicolumn{2}{c}{sign test} & \multicolumn{2}{c}{signed-rank test} \\
  $M_1$ (o)    &  $M_2$ ($+$) & $*$o & $*+$       & $*$o & $*+$     & $*$o  & $*+$    \\ \hline
   $D_1^2$     &  EA          &   3  &   {\bf 46} & 2    & {\bf 23} & 3     & {\bf 37}         \\
   $D_1^2$     &  K-V         &   18 &   {\bf 51} & 17   & {\bf 50} & 17    & {\bf 49}        \\
    K-V        &  EA          &   28 &   {\bf 53} & 25   & {\bf 49} & 29    & {\bf 51}        \\
\hline 
\multicolumn{8}{l}{see Fig. \ref{p-t-test-fig} and \S\ref{large-scale-data-results-section} for details}
\end{tabular}
\end{table}

Now lets summarise the results found so far, Table~\ref{result3-table} contains the results of the analysis of these
data sets over all time delay estimates regardless of noise and gap
size. Each row corresponds to a different statistic: 95\% CI, MSE, AE,
$\hat{\mu}$ and $\hat{\sigma}$. The accuracy is
measured by MSE and AE, where the best results are for K-V, which is
followed by EA. Since $\hat{\mu}$ and $\hat{\sigma}$ are the mean and
standard deviation over all estimates, these statistics can be seen as
a measurement of bias and variance of time delay estimates over all
data sets. Since the true delay is $\mu_0=5$, the minimum bias is for
$D_1^2$ ($|5.013-\mu_0|=0.013$); in second place is EA
($|5.015-\mu_0|=0.015$). The minimum variance is for K-V (0.68); in
second position is EA (0.79).

However, in practice, the noise is about 0.01 mag ($<0.106\%$) for the
optical data. In Tables~\ref{result3-1-table} to
\ref{result3-3-table}, and Figs.~\ref{t-test-fig}, \ref{MSE-fig} and \ref{p-t-test-fig},
the results are grouped by noise level. Tables~\ref{result3-1-table} and \ref{result3-2-table}
suggest that the results from EA are more statistically significant
than others (see \S \ref{large-scale-data-results-section}). 
Table~\ref{result3-3-table} shows that the results are also better from EA,
particularly when the noise is less than 0.106\%. If the noise level
is equal to 0.106\%, depending on the statistic, the best performance
is by either K-V or EA. When the noise is high (0.466\%), the best
results are from K-V. This can be also seen in Figs.~\ref{MSE-fig} and \ref{p-t-test-fig}.

The results from paired tests in Table \ref{result3-4-table} also suggest that the EA is capable of producing significantly superior time delay estimates, when compared to $D_1^2$ and K-V.

\subsubsection{PRH Data}\label{PRH-results-section}
Now lets compare the performance of the proposed method EA with another kind of data.
Here we use the artificial data generated by the PRH methodology (see \S 5.2 in
\cite{Press:1992:TTD} and \S6 in \cite{Cuevas:2006:HAT}),
so we compare only the PRH method against EA.  In fact, we compare the
performance of EA with the PRH method by fixing the PRH parameters to
those values used to generate the data (the ideal scenario). The
structure function (SF) to define the covariance matrix in the PRH
method is used in two ways: SF is fixed to its true value (SF*) and
then estimated following the PRH method (SF+).

Note that for these data there are several true time delays, $\mu_0$. 
In all cases, we use bounds\footnote{These bounds are also used to
estimate the structure function SF+ by the PRH method.} of $\mu_0 \pm
30$ days with unitary increments during the time delay analysis. The
measurement error is also fixed at its true value (variance of $1
\times 10^{-7}$) for all methods.

The results for the PRH method, SF* case, are in 
Table~\ref{result4-table}. The column $\mu_0$ denotes the true time delay,
which is also our hypothesis in the $t$-test, whereas $\eta=100$. The
following columns are the statistics used in this analysis (see section 
\S\ref{large-scale-data-results-section}). The last row is the average (Avg). The results for the SF+
case and for EA are in Table~\ref{result5-table} and Table~\ref{result6-table}
respectively. An analysis of bias ($|\mu_0-\hat{\mu}|$) and variance ($\hat{\sigma}$) is in Table~\ref{result7-table}.

\begin{table}
\renewcommand{\arraystretch}{1.3}
\caption{PRH data results: statistical analysis of the idealised method, PRH method with SF*.}
\label{result4-table}
 \begin{tabular}{c c c c c c c r}
 \hline 
  $\mu_0$     & ${\cal P}$ & ${\cal T}$ & 95\% CI  & CI range &  AE & MSE \\ \hline
     34     & 1.000    & 0.000  & 33.5 - 34.4 &  0.92    &  0.44 & 5.28  \\
     43     & 0.702    & 0.382  & 42.2 - 44.1 &  1.87    &  1.54 & 21.96 \\
     49     & 0.839    & 2.375  & 49.2 - 52.0 &  2.81    &  2.36 & 52.32 \\
     59     & 0.447    & 1.899  & 58.9 - 61.1 &  2.21    &  1.78 & 31.96 \\
     66     & 0.465    & 0.272  & 65.5 - 66.5 &  1.02    &  0.71 & 6.55  \\
     76     & 0.671    & 2.026  & 76.0 - 77.5 &  1.55    &  0.95 & 15.67 \\
     99     & 0.001    & 2.701  & 99.5 - 102.3&  2.86    &  2.79 & 55.39 \\ \hline
    Avg     & 0.374    &        &             &  1.89    &  1.51 & 27.02 \\
\hline 
\end{tabular}
\end{table}

\begin{table}
\renewcommand{\arraystretch}{1.3}
\caption{PRH data results: statistical analysis of PRH method with SF+.}
\label{result5-table}
 \begin{tabular}{c c c c c c c r}
 \hline 
  $\mu_0$    & ${\cal P}$  & ${\cal T}$ & 95\% CI     & CI range &  AE & MSE \\ \hline
     34    & 0.000    & -4.881 & 18.9 -27.6  &  8.72    & 22.79 & 593.45 \\
     43    & 0.210    & 1.260  & 81.8 - 48.2 &  6.46    & 13.51 & 266.19 \\
     49    & 0.315    & -1.008 & 43.7 - 50.7 &  6.96    & 15.15 & 307.95 \\
     59    & 0.977    & -0.028 & 54.6 - 63.1 &  8.51    & 19.66 & 455.20 \\
     66    & 0.257    &-1.138  & 58.7 - 67.9 &  9.23    & 22.21 & 542.99 \\
     76    & 0.031    & -2.188 & 66.7 - 75.5 &  8.83    & 20.95 & 513.97 \\
     99    & 0.407    & -0.832 & 93.0 - 101.4&  8.39    & 18.84 & 446.00 \\ \hline
    Avg    & 0.314    &        &             &  8.16    & 19.02 & 446.54 \\
\hline 
\end{tabular}
\end{table}

\begin{table}
\renewcommand{\arraystretch}{1.3}
\caption{PRH data results: statistical analysis of EA.}
\label{result6-table}
 \begin{tabular}{c c c c c c c r}
 \hline 
  $\mu_0$    & ${\cal P}$ & ${\cal T}$ & 95\% CI     & CI range &  AE & MSE \\ \hline
     34    & 0.600    & 0.525  & 32.3 - 36.8 & 4.58     & 7.68  & 132.27 \\
     43    & 0.330    & 0.977  & 42.5 - 44.4 & 1.96     & 2.28  &  24.34 \\
     49    & 0.389    & -0.864 & 46.8 - 49.8 & 2.94     & 3.99  &  54.69 \\
     59    & 0.684    & 0.407  & 57.2 - 61.9 & 4.35     & 7.28  & 119.00 \\
     66    & 0.957    &-0.052  & 63.9 - 67.9 & 3.98     & 7.16  &  99.53 \\
     76    & 0.301    &-1.039  & 73.0 - 76.9 & 3.83     & 6.89  &  93.42 \\
     99    & 0.830    & 0.214  & 96.7 - 101.7& 4.94     & 8.61  & 153.43 \\ \hline
    Avg    & 0.585    &        &             & 3.80     & 6.27  & 96.67  \\
\hline 
\end{tabular}
\end{table}

\begin{table*}
\renewcommand{\arraystretch}{1.3}
\caption{PRH data results: Bias ($|\mu_0 -\hat{\mu}|$) versus Variance ($\hat{\sigma}$).}
\label{result7-table}
 \begin{tabular}{c  c c c  c c c  c c c} 
 \hline 
           &  \multicolumn{3}{c}{PRH with SF*} & \multicolumn{3}{c}{PRH with SF+} & \multicolumn{3}{c}{EA} \\ 
  $\mu_0$  & $\hat{\mu}$ & $\hat{\sigma}$ & Bias & $\hat{\mu}$ & $\hat{\sigma}$ & Bias & $\hat{\mu}$ & $\hat{\sigma}$ &  Bias \\ \hline
     34    &   34.0      &{\bf 2.3 }      &{\bf 0.00}&   23.2      &  21.9          & 10.73&  34.4       & 11.8           & 0.46 \\
     43    &   43.1      &{\bf 4.7 }      &{\bf 0.18}&   45.0      &  16.2          & 2.05 &  43.7       &  4.8           & 0.79 \\
     49    &   50.6      &{\bf 7.0 }      &  1.68    &   47.2      &  17.5          & 1.77 &  48.4       &  7.7           &{\bf 0.57} \\
     59    &   60.0      &{\bf 5.5 }      &{\bf 0.07}&   58.9      &  21.4          & 0.06 &  59.9       &  9.7           & 0.96 \\
     66    &   66.0      &{\bf 2.5 }      &{\bf 0.07}&   63.3      &  23.2          & 2.65 &  65.7       &  9.2           & 0.21 \\
     76    &   76.7      &{\bf 3.8 }      &{\bf 0.79}&   71.1      &  22.2          & 4.87 &  75.1       &  10.2          & 0.86 \\
     99    &   100.9     &{\bf 7.2 }      &  1.95    &   97.2      &  21.1          & 1.76 &  99.8       &  12.2          &{\bf 0.89} \\ \hline
    Avg    &             &{\bf 4.7 }      &  0.81    &             &  20.5          & 3.41 &             &  9.4           &{\bf 0.68} \\
\hline 
\end{tabular}
\end{table*}

Results from paired tests are in Table \ref{result7-1-table}, where the p-values are shown. The best values are in bold, i.e., ${\cal P}< 0.05$.

\begin{table*}
\renewcommand{\arraystretch}{1.3}
\caption{PRH Data results: paired tests}
\label{result7-1-table}
 \begin{tabular}{c c c c c c c}
 \hline 
 &   \multicolumn{2}{c}{paired $t$-test} & \multicolumn{2}{c}{paired sign test} & \multicolumn{2}{c}{paired signed-rank test} \\
  $\mu_0$      &  SF* \& EA       & SF+ \& EA        &   SF* \& EA & SF+ \& EA  &  SF* \& EA   & SF+ \& EA   \\ \hline
    34         &  0.685           & {\bf 2.7e-5}$^a$ &  0.483          & {\bf 0.001}$^a$& 0.556          & {\bf 4.2e-5}$^a$ \\
    43         &  0.364           & 0.476            &  0.057          & 0.089          & {\bf 0.004}$^b$& 0.249        \\
    49         &  {\bf 0.021}$^a$ & 0.567            &  {\bf 0.012}$^a$& 0.193          & {\bf 0.002}$^a$& 0.425        \\
    59         &  0.919           & 0.673            &  {\bf 0.012}$^b$& 0.069          & 0.284          & 0.847        \\
    66         &  0.759           & 0.358            &  0.617          & 0.483          & 0.706          & 0.358        \\
    76         &  0.112           & 0.113            &  {\bf5e-4}$^b$  & 0.368          & {\bf 0.001}$^b$& 0.145        \\
    99         &  0.457           & 0.288            &  0.920          & {\bf 0.012}$^a$& 0.920          & 0.238        \\  
\hline 
\multicolumn{7}{l}{see \S\ref{PRH-results-section} for details.} \\
\multicolumn{7}{l}{$^a$ EA with minimum MSE.}  \\
\multicolumn{7}{l}{$^b$ PRH method with SF* has the minimum MSE.}
\end{tabular}
\end{table*}

Tables \ref{result4-table}-\ref{result7-table} show the results 
of the analysis applied to PRH data. Two versions of the PRH method were used: SF* and SF+ (see
\ref{PRH-results-section}). Here, contrary to the analysis of the large scale data, the noise level and gap size are
fixed (see \S \ref{data-section}). Rather, we use different short delays in the
range of 30 -- 100 days.

Comparing Tables~\ref{result4-table}-\ref{result6-table}, the highest
significance level (${\cal P}$) on $\mu_0=\{34,43\}$ is for the PRH method
(SF*). EA has better significance levels than the PRH method for
$\mu_0=\{66,99\}$, even considering the idealised case. For CI range,
MSE and AE statistics, SF* has the best performance on any
$\mu_0$. But, on all statistics EA performs better than
SF+. Therefore, EA is more accurate than PRH method (SF+).

In Table~\ref{result7-table}, we have $\hat{\mu}$ and $\hat{\sigma}$
as a time delay estimator and a measurement of uncertainty,
respectively. Hence, the column Bias is given by $|\mu_0 -
\hat{\mu}|$. The last row has the average (Avg). Therefore, the
minimum bias is for EA (on average). If we use $\hat{\sigma}$ as
variance measurement, the minimum variance is for PRH method
(SF*). But, EA has lower variance than SF+.

Table \ref{result7-1-table} shows that in few cases the paired difference between estimates is statistically significant (in bold). In those cases, EA is more accurate than PRH method (SF+), and in some cases EA is also more accurate than PRH method (SF*).

\subsubsection{Wiener Data}
Similarly, we compare the performance of EA with another kind of data
(see \S \ref{data-section}). These data come from a Bayesian approach
\cite{Harva:2006:IEEE}. We performed an analysis as above with
$\eta=225$, so the hypothesis to test is $H_0:\mu_0=35$.

The results\footnote{$\hat{\mu}$ and $\hat{\sigma}$ for the Bayesian
method are not reported in detail in \cite{Harva:2006:IEEE, HarvaR08}} 
are shown in Table~\ref{result8-table}. Since the data
were generated by the Bayesian method, we aim to compare such a method
with EA only.

\begin{table}
\renewcommand{\arraystretch}{1.3}
\caption{Wiener Data results: Statistical Analysis}
\label{result8-table}
 \begin{tabular}{c c c c c}
 \hline 
                &                  & \multicolumn{3}{c}{Data Set}      \\
  Method        & Statistic        & 0.1       & 0.2       & 0.4       \\ \hline
                & ${\cal P}$       & 0.59      & 0.63      &{\bf 0.06} \\
Bayesian method & ${\cal T}$       & 0.52      & 0.48      &{\bf 1.87} \\
                &  MSE             & 32.18     &{\bf 9.43} &{\bf 41.89} \\
                &  AE              & 1.84      &{\bf 1.94} &{\bf 3.7}  \\
                & $\hat{\mu}$      & 35.2      & 35.1      &{\bf 35.8} \\
                & $\hat{\sigma}$   & 5.7       &{\bf 3.1}  &{\bf  6.4} \\ \hline
                & ${\cal P}$       &{\bf 0.92} &{\bf 0.76} & 0.007\\
EA              & ${\cal T}$       &{\bf 0.09} &{\bf 0.30} & 2.70 \\
                &  MSE             &{\bf 10.06}& 23.25     & 66.99 \\
                &  AE              &{\bf 1.76} & 3.28      & 5.72  \\
                & $\hat{\mu}$      &{\bf 35.0} & 35.1      & 36.4 \\
                & $\hat{\sigma}$   &{\bf 3.1}  &  4.8      & 8.0  \\
\hline 
\end{tabular}
\end{table}

In Table \ref{result9-table} are the results from paired tests between Bayesian estimation method and our EA. We highlight in bold the p-values that are less than 0.05.

\begin{table}
\caption{Wiener Data results: Paired Tests}
\label{result9-table}
 \begin{tabular}{c c c c}
 \hline 
                & \multicolumn{3}{c}{Data Set}       \\
  Paired test   & 0.1       & 0.2       & 0.4        \\ \hline
  $t$-test      & 0.633     & 0.982     & 0.264      \\
  sign          & 0.505     & 0.893     &{\bf 0.007} \\
  signed-rank   & 0.827     & 0.766     &{\bf 0.005} \\
\hline 
\end{tabular}
\end{table}

In Table~\ref{result8-table}, the EA results are more significant for
data with noise levels of 0.1 and 0.2, where ${\cal P}$ is 0.92 and
0.76 respectively. But for a noise level of 0.4, it does not perform
as well.  In terms of bias, EA performs better for the data with a
noise level of 0.1, and ties in the case of that with 0.2 data and on
the noise of 0.4 data, the performance is not good enough. Talking
about variance $\hat{\sigma}$, EA is better on data set of noise 0.1
only. It needs to be emphasised though, that these data was generated by the Bayesian estimation method, 
so the comparison is positively biased towards the Bayesian method. However, Table \ref{result9-table} suggests that most of the paired 
differences between estimates by both methods are not statistically significant.

The poorer performance of the Bayesian method for low noise data,
compared to medium noise is explained by the posterior sampler, which
does not converge properly in some of the cases, giving biased
estimates. This can be easily avoided when analysing any given data
set, since the convergence can be assessed and the sampler re-run with
different parameters. With the repeated runs performed here, the same
parameters for the sampler with all of the datasets were used. In the
low noise case, the convergence measurement indicates that, for a
number of cases, these have not been optimal.

\subsubsection{Loss Functions}
Talking about loss functions, we compared the K-V and EA methods on Large Scale data (see
\S\ref{large-scale-data-results-section}). On the one hand, K-V employs
the negative log-likelihood ($Q$)\cite{Cuevas:2006:HAT}
as loss function, where the parameters $k$ and $\lambda$ are estimated via cross-validation
for trial time delays in the range $0\! -\!  10$. On the other hand, the
fitness function of EA is the MSE$_{CV}$ given by cross-validation.
We also compared the K-V method with EA 
by using the mean squared error\footnote{We point out that this mean squared error is different to MSE$_{CV}$; i.e., Eq. (\ref{log-loss-function}) without $\sigma_A^2(t_i)$ and $\sigma_B^2(t_i)$} as
the measurement of goodness of fit for K-V, rather than the
negative log-likelihood. Thus, the observational error is considered constant,
because the negative log likelihood fitting criterion in K-V has the form \cite{Cuevas:2006:HAT},
\begin{eqnarray}\label{log-loss-function}
\ Q = \sum_{i=1}^n \left ( \frac{(x_A(t_i)-h_A(t_i))^2}{\sigma_A^2(t_i)} + \frac{(x_B(t_i)-M \ominus h_B(t_i))^2}{\sigma_B^2(t_i)} \right ).
\end{eqnarray}
\noindent 
For the K-V method, the negative log-likelihood cost function (\ref{log-loss-function}) lead to more accurate estimates.

Furthermore, we tested another fitness function on these data. Instead of
MSE$_{CV}$, the fitness is given by the log-likelihood $Q$, where no
cross-validation is performed, since $k$ and $\theta$ are also
evolved. The results are that MSE$_{CV}$ performs better than $Q$,
where $Q$ is less time-consuming.

\subsubsection{Evolving Weights}
We also tried to evolve all the free parameters; that is, the weights
$\vec{\alpha}$ in (\ref{hA})-(\ref{hB}). This allowed us to avoid SVD
in (\ref{inverse}), which is $O(n^3)$ (without
cross-validation). However, the performance is poor because the number
of parameters to evolve increases with the number of samples $n$. In
fact, we tested this approach on artificial data 
and the performance was inferior to
that of our EA. 
This is due to an overwhelming number of variables.
Typically, using evolutionary approaches, one can well optimise in about 30-dimensional
search space
\cite{Tu:2004:StGA}. Perhaps a new framework can overcome this
problem.

\section{Conclusions}\label{conclusion-section}

From observed optical monitoring data, we suggest 419.6 days
as the best 
plausible value of the time delay between the two main images
(A and B) of the distant lensed quasar Q0957+561.
 
Regarding the artificial large scale data, the results from EA are important, because its accuracy in this
application is matched to the precision and low levels of noise with
which the current state-of-the art optical monitoring data are being
acquired for multiple-image time delay measurement
\cite{Colley:2003:ATC,Ovaldsen:2003:NAP,Pindor:2005:DGL,Eigenbrod:2005:TCM}.

From PRH data,  we conclude that EA is more accurate than PRH method (SF+), and
competitive with the idealised case (SF*). Unfortunately for the PRH
method, the idealised case does not exist in practice.

For the Wiener data, the conclusion is that for these data, EA
outperforms Bayesian estimation method for low levels of noise (0.1
data). In other cases, depending on the statistic, EA can show better,
equal or worse results.  We stress that the current real optical obsrevational
data are indeed characterised by low levels of observational noise.
It is exactly in this context where our EA method outperforms the
Bayesian estimation method.

An evolutionary algorithm has been introduced to form a novel
hybridisation with our kernel method, which is an automatic method to estimate
the time delay, kernels width, and SVD regularisation parameters. It is a successful application of EA driven by a model based
formulation to a real-world problem. The study of the statistical significance of results on different data shows that EA is a promising approach
on gravitational lensing, in astrophysics, but it can also be applied in other disciplines involving similar time series data.

\section{Future Work}\label{future-section}
One of the main issues to deal with, in future extensions
of this work, is speeding up our EA, which potentially
can be done in several ways.
First, we would seek the optimum procedure to invert
(\ref{inverse}) and its regularisation to deal with ill-conditioning (see
\ref{kernel-section}). Alternative ways for model selection, 
rather than using cross-validation, which is $O(n^3)$, are desirable (see in
\S\ref{evo-section}). The natural parallelisation of EA is
another research line to follow; e.g., see \cite{Cantu-Paz:2000:PGA}. 
To speed up the EA, fitness approximation techniques may also help with this problem \cite{Regis:2004:LFA,Paenke:2006:ESR}
since our fitness function is costly. Moreover, other evolutionary approaches in presence of 
uncertainties have not been tested \cite{Jin:2005:EOU,Arnold:2006:GNM}.

The time delay problem is still a huge issue in astrophysics. The next generation of large-scale monitoring projects
with dedicated telescopes, LSST (http://www.lsst.org/) and PAN-STARRS
(http://pan-starrs.ifa.hawaii.edu/) will produce monitoring data for thousands of potential multiply-imaged sources (as opposed to the 10 sources
for which data are now available). These datasets will be available in a few
years’ time, and a large effort is underway to develop algorithms that are far
more sophisticated than the currently available ones that would deal with
this data in an automated way, and can cope with the inevitable noise and
gap features of the data. We are attempting to develop robust methods that
will do this.

\ack{J.C. Cuevas-Tello would like to thank the sponsors at UASLP: PROMEP and FAI; grants C07-FAI-11-20.56 and PROMEP/103.5/08/1696, respectively. We also thank the reviewers who have helped to improve the paper significantly.
}

\bibliographystyle{elsart-num-sort}
\bibliography{ref_rev}

\end{document}